\documentclass[12pt,english]{extarticle}
\usepackage{newcent}
\usepackage[T1]{fontenc}
\usepackage[utf8]{inputenc}
\usepackage{geometry}
\usepackage{color}
\definecolor{page_backgroundcolor}{rgb}{1, 0.996094, 0.78125}
\pagecolor{page_backgroundcolor}
\usepackage{babel}
\usepackage{array}
\usepackage{prettyref}
\usepackage{float}
\usepackage{units}
\usepackage{multirow}
\usepackage{amsmath}
\usepackage{amsthm}
\usepackage{amssymb}
\usepackage{graphicx}
\usepackage{esint}
\usepackage[unicode=true,pdfusetitle,
 bookmarks=true,bookmarksnumbered=false,bookmarksopen=false,
 breaklinks=false,pdfborder={0 0 1},backref=section,colorlinks=true]
 {hyperref}

\makeatletter

\providecommand{\tabularnewline}{\\}

\numberwithin{equation}{section}
\numberwithin{figure}{section}

\usepackage{caption}
\usepackage[section]{placeins}
\usepackage{float}
\usepackage{graphicx}
\usepackage{epigraph}

\usepackage{chngcntr}
\counterwithout{figure}{section} 

\counterwithout{equation}{section} 

\usepackage{eso-pic}

\usepackage{fancyhdr}
\date{}

\definecolor{olivegreen}{RGB}{0,175,0}

\hypersetup{colorlinks=true,citecolor=olivegreen,linkcolor=blue, urlcolor=blue}

\makeatother

\begin{document}
\pagenumbering{gobble} 

\title{Eigenvalue and Eigenfunction for the $PT$-symmetric Potential $V=-\left(ix\right)^{N}$}

\maketitle
\vspace{-1.6cm}

\begin{center}
\author{\large Cheng Tang$^1$ \; \normalsize and \large \; Andrei Frolov$^2$\\[0.1in] 
\normalsize\it Department of Physics, \; Simon Fraser University\\[0.1cm]
V5A 1S6, \; Burnaby, \; BC, \; Canada\\[0.1cm]
$^1$\href{mailto:cta63@sfu.ca}{cta63@sfu.ca} \quad
\and 
$^2$\href{mailto:frolov@sfu.ca}{frolov@sfu.ca} }
\par\end{center}

\begin{center}
\date{February 27, 2017}
\par\end{center}

\smallskip{}

\begin{abstract}
\cfoot{\large\bf\thepage}

If replace the Hermiticity from conventional quantum mechanics with
the physically transparent condition of parity-time reflection symmetry
($PT$-symmetry), the non-Hermitian Hamiltonian still guarantees that
its entire energy spectrum is real if the Hamiltonian has unbroken
$PT$-symmetry. If its $PT$-symmetry is broken, then two cases can
happen - its entire energy spectrum is complex for the first case,
or a finite number of real energy levels can still be obtained for
the second case. This was ``officially'' discovered on a paper by
Bender and Boettcher since 1998 when the energy spectrum from the
$PT$-symmetric Hamiltonian $\mathcal{H}=p^{2}-\left(ix\right)^{N}$
with $x\in\mathbb{C}$ was examined within one pair of Stokes wedges. 

To better understanding differential equation in complex plane, for
this Hamiltonian we discuss the following three questions in this
paper. First, since their paper used a Runge-Kutta method to integrate
along a path at the center of the Stokes wedges to calculate eigenvalues
$E$ with high accuracy, we wonder if the same eigenvalues can be
obtained if integrate along some other paths in different shapes.
Second, what the corresponding eigenfunctions look like? Should the
eigenfunctions be independent from the shapes of path or not? Third,
since for large $N$ the Hamiltonian contains many pairs of Stokes
wedges symmetric with respect to the imaginary axis of $x$, thus
multiple families of real energy spectrum can be obtained. What do
they look like? Any relation among them? 
\end{abstract}
\cfoot{\thepage}

\tableofcontents{}


\pagenumbering{arabic}
\cfoot{\large\bf\thepage}

\section{Introduction}

In quantum mechanics, the sign of position operator $\hat{x}$ and
the momentum operator $\hat{p}$ can be changed by the parity reflection
operator $P$ in the following way\cite{Make_sense}:

\begin{equation}
P\hat{x}P=-\hat{x}\qquad P\hat{p}P=-\hat{p}\qquad PiP=i\text{ ,}\label{eq:parity_operator}
\end{equation}
where, however, the sign of the complex number $i$ is unchanged.
If we apply the time reversal operator $\hat{T}$ instead, then
\begin{equation}
T\hat{x}T=\hat{x}\qquad T\hat{p}T=-\hat{p}\qquad TiT=-i\text{ ,}\label{eq:time_operator}
\end{equation}
where the sign of the momentum operator and of the complex number
are changed. We say a Hamiltonian $\mathcal{H}$ is $PT$-symmetric
if the combined operator $PT$ commutes with $\hat{\mathcal{H}}$
such that 
\begin{equation}
\left[PT,\mathcal{\hat{\mathcal{H}}}\right]=\mathcal{\hat{\mathcal{H}}}\left(PT\right)-\left(PT\right)\mathcal{\hat{\mathcal{H}}}=0\text{ .}\label{eq:H_and_PT_commute}
\end{equation}
For example, the Hamiltonian $\mathcal{H}=p^{2}-\left(ix\right)^{N}$
with $x\in\mathbb{C}$ and $N\in\mathbb{R}$ is $PT$-symmetric.

The discovery\cite{Real_Spectra,dorey2001spectral} that an entirely
real energy spectrum can be obtained from the non-Hermitian but $PT$-symmetric
Hamiltonian was a surprise to scientific communities in 1998. Since
then, the developments in $PT$-symmetric quantum theory rapidly grew
- at least 50 experiments to observe $PT$-symmetric system were published
during the last 10 years. Those experiments told us that it was possible
to experimentally measure complex eigenvalue, and observe broken and
unbroken $PT$-symmetry. 

It's well-known\cite{Make_sense} that within a specific pair of Stokes
wedges the $PT$-symmetry of $\mathcal{H}=p^{2}-\left(ix\right)^{N}$
is unbroken if $N\geq2$, so that the entire energy spectrum is positive;
and broken if $N<2$, so that only a finite number of positive energy
levels can be found. However, this conclusion is not satisfied for
the curiosity of any enthusiastic student. As far as we know, most
students from physical science are unfamiliar with the concept of
Stokes wedge. Driven by curiosity, they would ask something similar
to the following questions: 
\begin{enumerate}
\item Since Bender and Boettcher\cite{Real_Spectra} used a Runge-Kutta
method to integrate along a path at the center of the Stokes wedges
to calculate eigenvalues $E$ with high accuracy, we wonder if the
same eigenvalues can be obtained if integrate along some other paths
in different shapes. In other words, are those eigenvalues independent
from the shape of path or not?
\item So far we have not seen any research about its eigenfunctions, and
do not know why the research of eigenfunction should be ignored. In
this paper, can we provide a detailed study to fill the gap?
\item Due to the existence of multiple pairs of Stokes wedges, the Hamiltonian
$\mathcal{H}=p^{2}-\left(ix\right)^{N}$ must contain multiple families
of real energy spectrum if $N$ is large enough. We are really curious
to see what the spectra look like.
\end{enumerate}
It's difficult to answer these three questions by using rigorous mathematics
which is beyond university-level and in fact we don't know. To answer
them pedagogically, we implement the strategy - ``seeing is believing''.
A lot of figures and tables are shown in this paper to help students
visualize data, Stokes wedges and the relation among them. How we
answer the questions is based on the observation of the data, rather
than rigorous mathematics. We believe empirical observation and conjecture
are the first and crucial step to deepen our understanding upon rather
complicated concept.

Our paper is organized in the following way. In Sec.\eqref{sec:Local-asymptotic-analysis},
the concept of Stokes wedges is introduced. In Sec.\eqref{sec:Numerical-approximation},
we show two ingredients needed for the numerical calculation of the
eigenvalue - algorithm and parametrization, then we provide answers
for the first two questions by discussing the results for the specific
values of $N$. In Sec.\eqref{sec:Multiple_families}, after introduced
the WKB approximation, we plot and discuss the first four families
of the spectrum. 

\section{\label{sec:Local-asymptotic-analysis}Local asymptotic analysis for
the potential $V=-\left(ix\right)^{N}$}

Consider $1$D Schrodinger equation in the complex plane with $N\in\mathbf{\mathbb{R}}$
and $x\in\mathbf{\mathbb{C}}$, 
\begin{equation}
-\psi^{\prime\prime}\left(x\right)-\left(ix\right)^{N}\psi\left(x\right)=E\psi\left(x\right)\text{,}\label{eq:Eigenvalue_eq}
\end{equation}
with boundary condition $\psi\left(x\right)\rightarrow0$ and $\psi^{\prime}\left(x\right)\rightarrow0$
as $|x|\rightarrow\infty$.

\begin{figure}[H]
\caption{\label{fig:many_wedges_multiplot}All Stokes wedges for non-negative
integer $N$ and all corresponding turning points (yellow point with
black edge). Although the locations of turning points are eigenvalue
$E$-dependent, for the purpose of visualization here we set $E$
as an appropriate constant.}

\centering{}\includegraphics[scale=1.38]{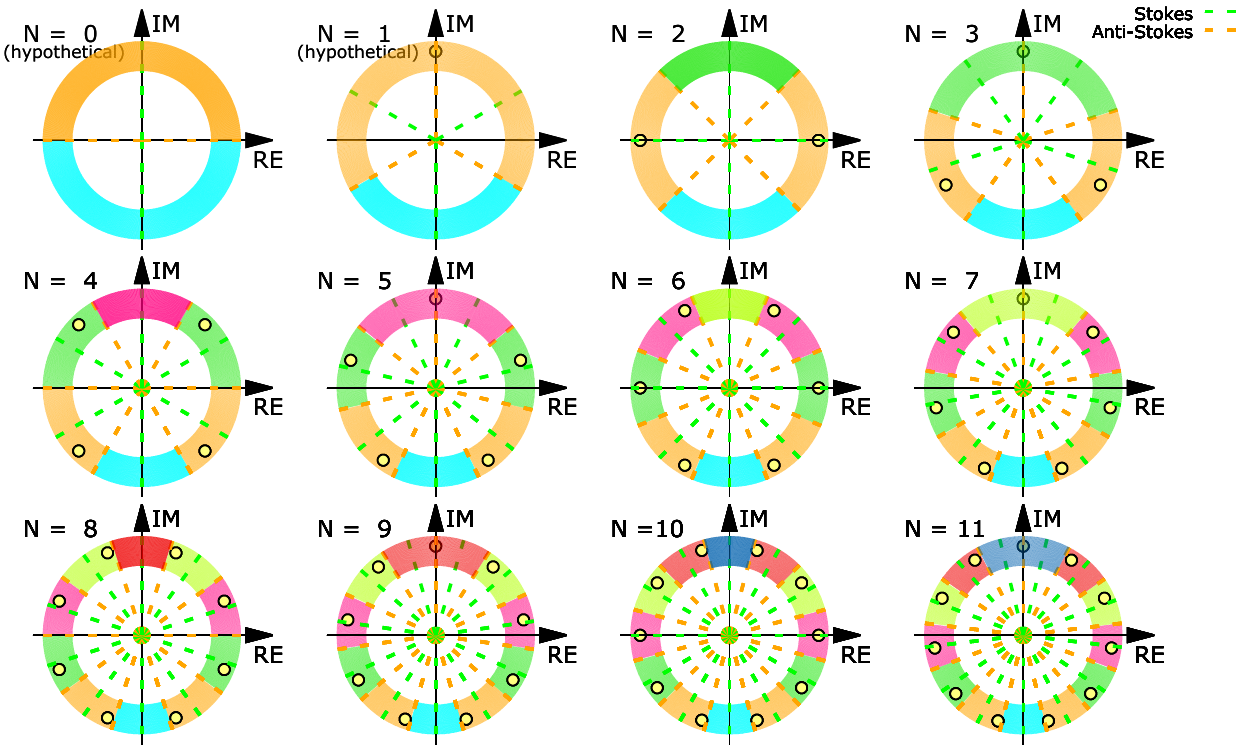}
\end{figure}

\begin{figure}[H]
\caption{\label{fig:chosen_wedges_multiplot}The chosen Stokes wedges for non-negative
integer $N$ and all corresponding turning points (yellow point with
black edge).}

\centering{}\includegraphics[scale=1.38]{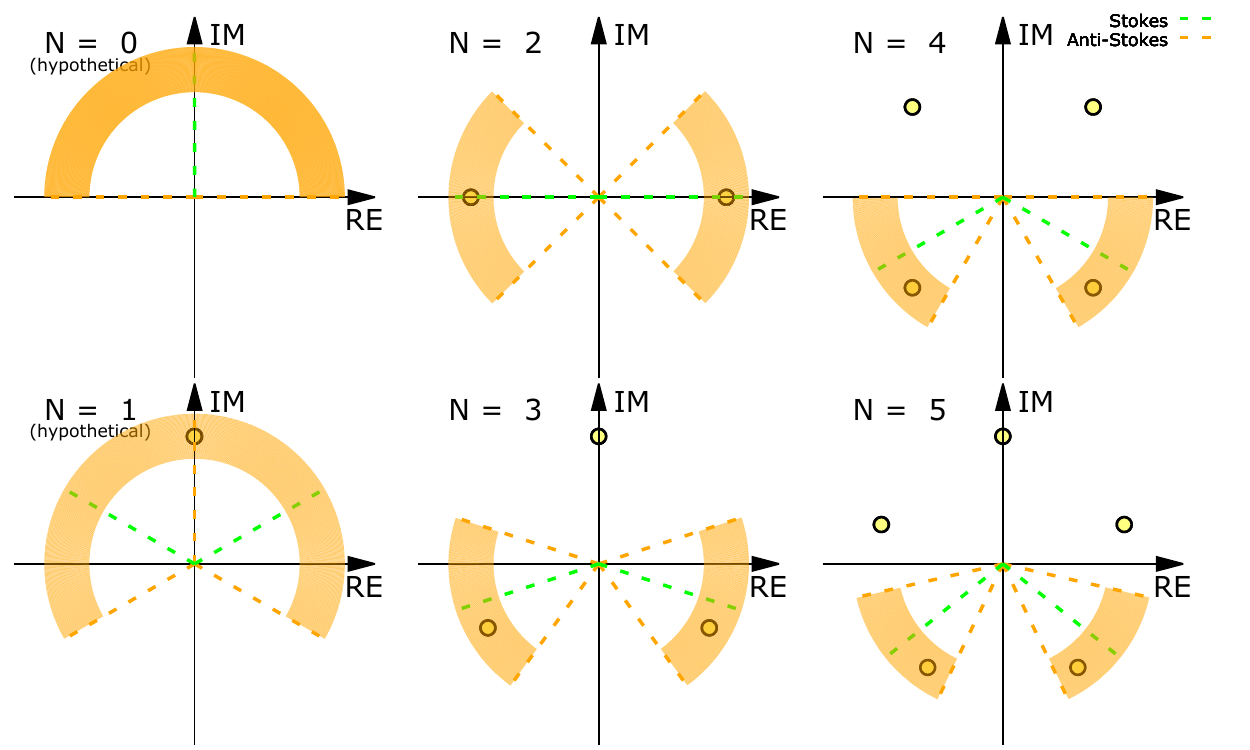}
\end{figure}

We guess that \prettyref{eq:Eigenvalue_eq} has a solution with the
form $\psi\left(x\right)=e^{S\left(x\right)}$ where $S\left(x\right)=ax^{b}$
with $a\in\mathbf{\mathbb{C}}$, $a\neq0$ and $b>0$. So we substitute
our ansatz into \prettyref{eq:Eigenvalue_eq}, neglect\cite{bender1999advanced}
those terms whose modulus are orders of magnitude smaller than the
rest when $\left|x\right|\rightarrow\infty$, and finally obtain the
following asymptotic relations as $\left|x\right|\rightarrow\infty$,
\begin{equation}
\psi\left(x\right)\sim C_{1}\left|x\right|^{-\frac{N}{4}}\exp\left[\pm\frac{i^{\frac{N}{2}+1}}{\frac{N}{2}+1}x^{\frac{N}{2}+1}\right]\qquad\mbox{when }N\geq2\text{;}\label{eq:leading_order_behavior_N_bigger_2}
\end{equation}

\begin{equation}
\psi\left(x\right)\sim C_{2}\left|x\right|^{-\frac{N}{4}}\exp\left[\pm\frac{i^{\frac{N}{2}+1}}{\frac{N}{2}+1}x^{\frac{N}{2}+1}\mp\frac{E\left(x^{-\frac{N}{2}+1}\right)}{\left(-N+2\right)i^{\frac{N}{2}+1}}\right]\qquad\mbox{when }\frac{2}{3}\leq N<2\text{;}\label{eq:N_less_2}
\end{equation}
\begin{align}
\psi\left(x\right) & \sim C_{3}\left|x\right|^{-\frac{N}{4}}\exp\left[\pm\frac{i^{\frac{N}{2}+1}}{\frac{N}{2}+1}x^{\frac{N}{2}+1}\mp\frac{E\left(x^{-\frac{N}{2}+1}\right)}{\left(-N+2\right)i^{\frac{N}{2}+1}}\pm\frac{E^{2}}{8i^{\frac{3N}{2}+1}\left(1-\frac{3N}{2}\right)}x^{-\frac{3N}{2}+1}\right]\label{eq:N_less_2over3}\\
 & \qquad\mbox{when }0<N<\frac{2}{3}\text{,}\nonumber 
\end{align}
where $C_{1}$, $C_{2}$ and $C_{3}$ are some constants. Note that
the asymptotic relation \prettyref{eq:leading_order_behavior_N_bigger_2}
for $N\geq2$ is independent from the eigenvalue $E$. To satisfy
the boundary condition, we expect that its dominant contribution in
the leading-order behavior vanishes such that 

\[
\exp\left[S_{1,2}\left(x\right)\right]\equiv\exp\left[\pm\frac{i^{\frac{N}{2}+1}}{\frac{N}{2}+1}x^{\frac{N}{2}+1}\right]\rightarrow0\qquad\mbox{ as }\left|x\right|\rightarrow\infty\text{.}
\]
$\exp\left[S_{1,2}\left(x\right)\right]$ approaches to zero in the
fastest speed if the oscillatory behavior of the exponential is zero,
in other words\cite{bender1999advanced}, 
\begin{equation}
\operatorname{Im}\left[S_{1}\left(x\right)-S_{2}\left(x\right)\right]=0\Longrightarrow\operatorname{Im}\left[i^{\frac{N}{2}+1}x^{\frac{N}{2}+1}\right]=0\text{,}\label{eq:Def_anti-Stokes}
\end{equation}
which is our definition of ``Stokes lines''. With $x=re^{i\theta}$,
\prettyref{eq:Def_anti-Stokes} becomes 
\[
\frac{\pi}{2}\left(\frac{N}{2}+1\right)+\theta\left(\frac{N}{2}+1\right)=\pm k\pi\qquad\mbox{for }k=0,1,2,3,\cdots\text{,}
\]
which yields
\begin{equation}
\begin{cases}
\theta_{left}=-\pi+\frac{N-4k+2}{N+2}\frac{\pi}{2}\\
\theta_{right}=-\frac{N-4k+2}{N+2}\frac{\pi}{2}
\end{cases}\mbox{ for }k=0,1,2,3,\cdots\text{,}\label{eq:many_anti_Stokes}
\end{equation}
so that
\begin{equation}
\theta_{left}=-\pi-\theta_{right}\text{.}\label{eq:Relation_left_and_right}
\end{equation}
We plot all those Stokes lines on Fig.\eqref{fig:many_wedges_multiplot}.
When $k=1$ we obtain 
\begin{equation}
\begin{cases}
\theta_{left}=-\pi+\frac{N-2}{N+2}\frac{\pi}{2}\\
\theta_{right}=-\frac{N-2}{N+2}\frac{\pi}{2}
\end{cases}\text{.}\label{eq:location_anti_Stokes_lines}
\end{equation}
These Stokes lines are plotted on Fig.\eqref{fig:chosen_wedges_multiplot}.

The locations of ``anti-Stokes lines'' are defined as 
\begin{equation}
\operatorname{Re}\left[S_{1}\left(x\right)-S_{2}\left(x\right)\right]=0\text{,}\label{eq:Def_Stokes_lines}
\end{equation}
when the exponential is purely oscillatory. Solving \prettyref{eq:Def_Stokes_lines}
yields 
\[
\theta_{1}=\frac{\pi}{N+2}-\frac{\pi}{2}+\frac{2\left(j-1\right)\pi}{N+2}\qquad\theta_{2}=\frac{\pi}{N+2}-\frac{\pi}{2}+\frac{2j\pi}{N+2}\qquad\mbox{for }j=0,1,2,3,\cdots\text{,}
\]
which define the width of the ``Stokes sector'' or ``Stokes wedge''
\begin{equation}
\triangle=\left|\theta_{1}-\theta_{2}\right|=\frac{2\pi}{N+2}\text{.}\label{eq:width_Stokes_Sector}
\end{equation}
The shape of a Stokes wedge is not really like a wedge or a slice
of pie. They are asymptotic concepts. The angular opening $\triangle$
from \prettyref{eq:width_Stokes_Sector} of the wedge only refers
to the opening for $\left|x\right|$ at certain range of complex infinity.

Except the top and bottom wedges which contain the imaginary axis,
Fig.\eqref{fig:many_wedges_multiplot} shows that the rest of all
wedges form pairs symmetric with respect to the imaginary axis. Each
pair are labeled with a color - orange, green, pink, yellow, red,
etc. We may sometimes call those pairs as ``$PT$-symmetric Stokes
wedges''. Each pair contains a pair of turning points which are also
symmetric with respect to the imaginary axis. The larger the $N$
is, the more pairs of wedges and of turning points are. Any wedge
whose anti-Stokes line coincides with the imaginary axis only shares
one singular turning point with its pair, and that singular turning
point must be located on the imaginary axis. On Fig.\eqref{fig:many_wedges_multiplot},
when $N=0$ and $N=1$, we label the wedges as ``hypothetical wedges'',
because the locations of these wedges on the figure are actually not
true for $N<2$ according to \prettyref{eq:N_less_2} and \prettyref{eq:N_less_2over3},
where these wedges are eigenvalue $E$-dependent.

Since different pair of wedges will pose different eigenvalue problem,
to proceed, we now only focus on one pair of wedges by choosing the
orange pair with the Stokes lines defined from \prettyref{eq:location_anti_Stokes_lines}
shown on Fig.\eqref{fig:chosen_wedges_multiplot} to calculate the
eigenvalue. 

\section{\label{sec:Numerical-approximation}Numerical approximation}

\subsection{Levenberg-Marquardt algorithm}

To solve the eigenvalue problem \prettyref{eq:Eigenvalue_eq}, by
Levenberg-Marquardt algorithm (LMA) we choose to minimize a square-function
$F\left(x_{\infty},E\right)$ of the following complex modulus with
respect to the energy $E$,

\[
F\left(x_{\infty},E\right)\equiv\left|f\left(x_{\infty},E\right)-\psi\left(x_{\infty},E\right)\right|^{2}
\]
where $x_{\infty}$ is the right boundary point located within a right
Stokes wedge and $f\left(x_{\infty},E\right)$ can be calculated by
Gauss-Legendre integration method\cite{Gauss_legendre} (GLI), an
implicit Runge-Kutta method. Due to the boundary condition $\lim_{\left|x\right|\rightarrow\infty}\psi\left(x\right)=0$,
\begin{equation}
F\left(x_{\infty},E\right)=\left|f\left(x_{\infty},E\right)\right|^{2}\text{.}\label{eq:F_expression}
\end{equation}
LMA is an iterative procedure, where the previous estimate $E$ is
replaced by a new estimate, $E+\delta E$, for each iterative step.
We can approximate $f\left(x_{\infty},E+\delta E\right)$ by 
\begin{equation}
f\left(x_{\infty},E+\delta E\right)\approx f\left(x_{\infty},E\right)+J\,\delta E\text{,}\label{eq:f_deltaE_expression}
\end{equation}
where $J$ is the gradient of $f\left(x_{\infty},E\right)$ with respect
to $E$, 
\[
J=\frac{\partial f\left(x_{\infty},E\right)}{\partial E}\text{.}
\]
In our case, $f\left(x_{\infty},E\right)$ is complex and so is $E$.
Let $u$, $v$, $a$ and $b$ be real such that 
\[
f\left(x_{\infty},E\right)=u\left(a,b\right)+iv\left(a,b\right)\qquad E=a+ib\text{,}
\]
then we have the following Jacobian matrix $\mathbf{J}$ 
\begin{equation}
\mathbf{J}=\left[\begin{array}{cc}
\frac{\partial u}{\partial a} & \frac{\partial u}{\partial b}\\
\frac{\partial v}{\partial a} & \frac{\partial v}{\partial b}
\end{array}\right]\text{,}\label{eq:Jacobian_matrix}
\end{equation}
and \prettyref{eq:f_deltaE_expression} in vector notation is 
\begin{equation}
\mathbf{f}\left(\mathbf{x}_{\infty},\mathbf{E}+\mathbf{\delta E}\right)\approx\mathbf{f\left(x_{\infty},E\right)}+\mathbf{J}\,\mathbf{\delta E}\text{,}\label{eq:f_deltaE_expression2}
\end{equation}
where 
\[
\mathbf{f\left(x_{\infty},E\right)}=\left[\begin{array}{c}
u\\
v
\end{array}\right]\qquad\mathbf{\delta E}=\left[\begin{array}{c}
\delta a\\
\delta b
\end{array}\right]\text{.}
\]
By \prettyref{eq:F_expression} and \prettyref{eq:f_deltaE_expression2},
we obtain 
\begin{equation}
F\left(\mathbf{x_{\infty},E}+\mathbf{\delta E}\right)=\left[\mathbf{f\left(x_{\infty},E\right)}+\mathbf{J}\,\mathbf{\delta E}\right]^{2}=\left[\mathbf{f\left(x_{\infty},E\right)}+\mathbf{J}\,\mathbf{\delta E}\right]^{T}\left[\mathbf{f\left(x_{\infty},E\right)}+\mathbf{J}\,\mathbf{\delta E}\right]\text{.}\label{eq:square_function}
\end{equation}
To find the minimum, we set 
\begin{align*}
\frac{\partial F\left(\mathbf{x_{\infty},E}+\mathbf{\delta E}\right)}{\partial\left(\mathbf{\delta E}\right)} & =0\text{.}
\end{align*}
Hence, 
\[
\mathbf{\delta E}=-\left(\mathbf{J}^{T}\mathbf{J}\right)^{-1}\mathbf{J}^{T}\mathbf{f\left(x_{\infty},E\right)}\text{.}
\]
Due to Levenberg's and Marquardt's modification on the last equation,
we have a damped factor $\lambda$, which is a positive parameter,
such that 
\begin{equation}
\mathbf{\delta E}\mathbf{=}-\left[\mathbf{J}^{T}\mathbf{J}+\lambda\mbox{diag}\mathbf{\left(\mathbf{J}^{T}\mathbf{J}\right)}\right]^{-1}\mathbf{J}^{T}\mathbf{f\left(x_{\infty},E\right)}\text{,}\label{eq:Levenberg_result}
\end{equation}
where $\mbox{diag}\mathbf{\left(\mathbf{J}^{T}\mathbf{J}\right)}$
means a diagonal matrix with entries on the diagonal from the matrix
$\mathbf{J}^{T}\mathbf{J}$. If the function $F\left(\mathbf{x_{\infty},E}_{new}\right)\leq F\left(\mathbf{x_{\infty},E}_{old}\right)$
after a single iterative step, we update $\mathbf{E}_{old}$ by 
\begin{equation}
\mathbf{E}_{new}=\mathbf{E}_{old}+\mathbf{\delta E}=\mathbf{E}_{old}-\left[\mathbf{J}^{T}\mathbf{J}+\lambda\mbox{diag}\mathbf{\left(\mathbf{J}^{T}\mathbf{J}\right)}\right]^{-1}\mathbf{J}^{T}\mathbf{f\left(x_{\infty},E\right)}\text{,}\label{eq:Update_E_result}
\end{equation}
and meanwhile decrease the value $\lambda_{old}$ by a factor, for
example, $\lambda_{new}=\nicefrac{\lambda_{{\scriptscriptstyle old}}}{\sqrt{2}}$.
If after a single iterative step $F\left(\mathbf{x_{\infty},E}_{new}\right)>F\left(\mathbf{x_{\infty},E}_{old}\right)$,
this means our $\lambda$ value is too small and we then increase
$\lambda_{old}$ by a factor, for example, $\lambda_{new}=10\lambda_{old}$.
And the eigenvalue will not be updated so that we still have 
\[
\mathbf{E}_{new}=\mathbf{E}_{old}\text{.}
\]
How we adjust the value of $\lambda$ becomes important to efficiently
find the eigenvalue, yet so far there is no absolutely best way to
optimize the value of $\lambda$. 

\subsection{Parametrization, eigenvalue and eigenfunction}

\begin{figure}[H]
\caption{\label{fig:bad_parametrize}The eigenfunction of the ground state
for $N=4$ along the Stokes lines (non-differentiable at origin) with
$r_{0}=4$. The vertical-blue dotlines represent two numerical infinities
$\pm\operatorname{Re}\left(x_{0}\right)=\pm r_{0}\cos30^{\circ}$.
Along the Stokes lines, the numerical result for the eigenvalue $E$
does not change even though the shape of the eigenfunction is not
smooth, in comparison with the smooth eigenfunctions associated with
the hyperbolic paths.}

\centering{}\includegraphics[scale=1.15]{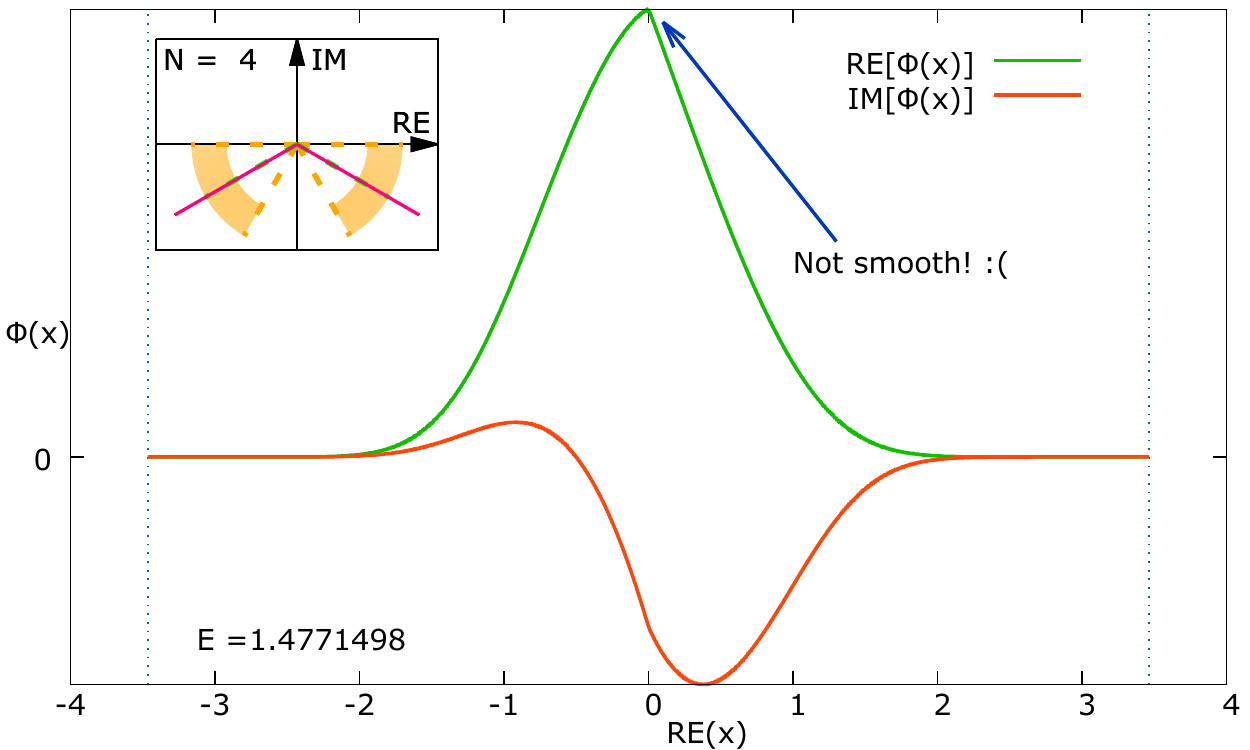}
\end{figure}

We set up the following initial condition at the numerical infinity
$x_{0}\equiv\infty_{left}$ within the left Stokes wedge on Fig.\eqref{fig:chosen_wedges_multiplot}:
\begin{equation}
x_{0}=r_{0}\exp\left(i\theta_{left}\right)\qquad\psi\left(x_{0}\right)=0\qquad\frac{d\psi\left(x_{0}\right)}{dx}=10^{-7}\text{,}\label{eq:initial_condition}
\end{equation}
where $\theta_{left}$ is defined by \prettyref{eq:location_anti_Stokes_lines}
and $r_{0}$ is the complex modulus of the numerical infinity $x_{0}$.
In other words, $r_{0}$ is the distance between the origin and the
point where the wave function and the derivative of the wave function
almost vanish. By observation, we set $r_{0}=4$. To have the fastest
convergence to the eigenvalue, we're tempted to use GLI to integrate
along the two Stokes lines given by \prettyref{eq:location_anti_Stokes_lines}.
However, they are connected at the origin where is non-differentiable
for $N\neq2$. Since this causes non-smoothness (See Fig.\eqref{fig:bad_parametrize})
on the eigenfunction at the origin, one way to have smooth-looking
eigenfunction is to integrate along some new paths, which should satisfy
the following four conditions:
\begin{enumerate}
\item The potential $-\left(ix\right)^{N}$ has a numerical cut on the positive-imaginary
axis. So the new path must not cross it; otherwise we must have a
different eigenvalue problem. 
\item The new path must go from one complex infinity within one Stokes wedge
and back to the other complex infinity in the other Stokes wedge.
These two complex infinities are symmetric with respect to the imaginary
axis of $x$.
\item The new path is smooth everywhere and can be parametrized by a differentiable
function.
\item Since GLI converges fastest if integrate along the two Stokes lines,
it would be more efficient if the path or the differentiable function
has two asymptotic lines coincident with the locations of the two
Stokes lines. 
\end{enumerate}

\begin{figure}[H]
\caption{\label{fig:bad_become_good}The eigenfunctions of the ground state
for $N=4$ along the differentiable (hyperbolic) paths with three
different values of $a$ defined by the hyperbolic equation $Y_{-}=-a\sqrt{1+\frac{X^{2}}{b^{2}}}$.
All three paths have the same eigenvalue. The vertical-blue dotlines
represent two numerical infinities $\pm\operatorname{Re}\left(x_{0}\right)=\pm r_{0}\cos30^{\circ}$
with $r_{0}=4$.}

\centering{}\includegraphics[scale=1.15]{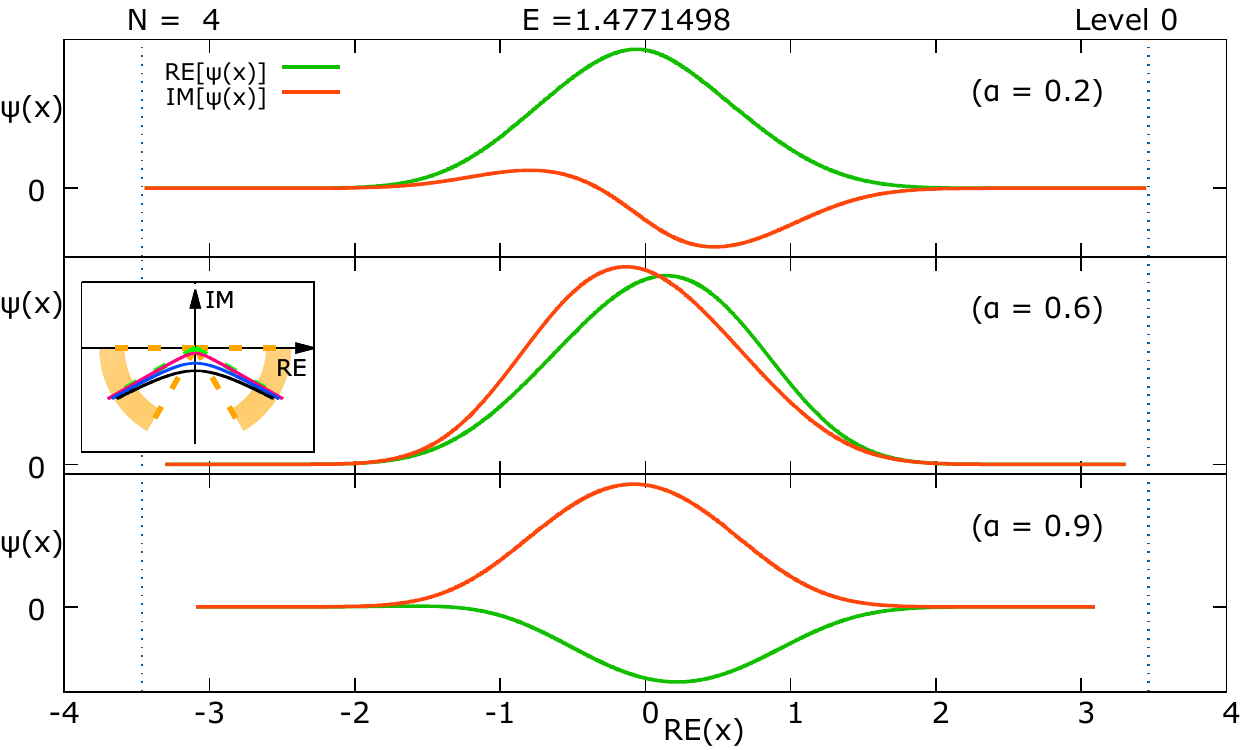}
\end{figure}

To satisfy all four conditions and since Fig.\eqref{fig:chosen_wedges_multiplot}
shows that the two Stokes lines move below the real axis when $N>2$,
the best differentiable function used for the parametrization when
$N>2$ must be hyperbola shown on the miniplot on Fig.\eqref{fig:bad_become_good}.
In this hyperbolic parametrization, we treat $\operatorname{Re}\left(x\right)$
as the parameter so that 
\begin{equation}
x=\operatorname{Re}\left(x\right)+i\operatorname{Im}\left(x\right)=\operatorname{Re}\left(x\right)-i\sqrt{a^{2}+\left[\operatorname{Re}\left(x\right)\right]^{2}\left(\tan\theta\right)^{2}}\text{,}\label{eq:hypola_parametriz_eq}
\end{equation}
where the angle $\theta$ between one of the asymptotic lines and
the horizontal axis is $\theta=\arctan\left(\frac{a}{b}\right)$.
Also, $\theta=\theta_{right}$ from \prettyref{eq:location_anti_Stokes_lines}.
Fig.\eqref{fig:bad_become_good} shows that the shapes of the corresponding
eigenfunctions are smooth and different-looking since the values of
$a$ defined by the hyperbolic equation $Y_{-}=-a\sqrt{1+\frac{X^{2}}{b^{2}}}$
is different. For the upcoming work, we choose $a=0.2$ since this
hyperbola is quite close to the location of the two Stokes lines and
meanwhile keeps the shape of eigenfunction smooth. 

\begin{figure}[H]
\caption{\label{fig:Nless2_function}The differentiable path for $N<2$ by
choosing $c=\frac{1}{10}$ and $t=8$. }

\centering{}\includegraphics[scale=0.55]{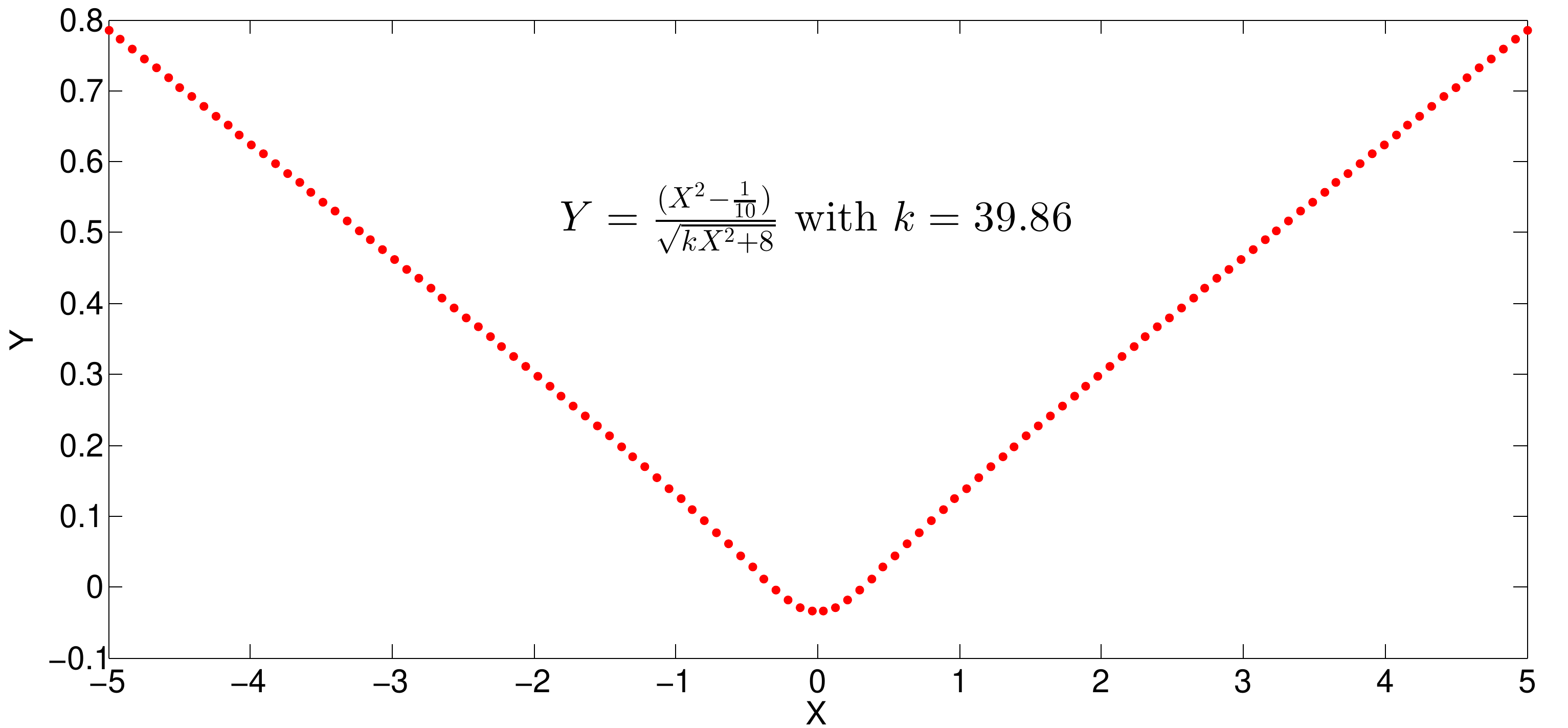}
\end{figure}

For $0<N<2$, Fig.\eqref{fig:chosen_wedges_multiplot} shows that
the two Stokes lines move above the real axis. Does the function satisfy
the four conditions exist? Yes. As $X\rightarrow\pm\infty$ the following
function with $k>0$, real parameters $c$ and $t$ 
\begin{equation}
f\left(X\right)=\frac{X^{2}-c}{\sqrt{kX^{2}+t}}\label{eq:function_for_parametrix_Nless2}
\end{equation}
has two asymptotic lines:
\[
Y=\frac{1}{\sqrt{k}}X\mbox{ \:\ as }X\rightarrow+\infty\text{,}\qquad Y=-\frac{1}{\sqrt{k}}X\mbox{ \:\ as }X\rightarrow-\infty\text{.}
\]
The angle $\theta$ between the asymptotic line associated with $X\rightarrow+\infty$
and the horizontal axis satisfies 
\[
k=\left(\frac{1}{\tan\theta}\right)^{2}\text{.}
\]
Fig.\eqref{fig:Nless2_function} shows a good news that the function
satisfies all four conditions.

With these parametrizations \prettyref{eq:hypola_parametriz_eq},
\prettyref{eq:function_for_parametrix_Nless2} and along the real
axis of $x$, we are able to use LMA and GLI to find all eigenvalues
and eigenfunctions associated with all pairs of wedges defined by
Fig.\eqref{fig:many_wedges_multiplot}. The final result of eigenvalues
is shown on Fig.\eqref{fig:final_result_eigenvalue_all_wedges}, and
we will explain it later.

\subsubsection{When $N=2$}

The concept of Stokes wedge implies that the same eigenvalue is obtained
if we integrate along different paths, as along as the conditions
1 and 2 are satisfied. However, can we justify this by numerical analysis?
How about their eigenfunctions? Are they independent from the shape
of path or not?

\begin{figure}[H]
\caption{\label{fig:indepdent_paths_N=00003D2}Four distinct contour paths
we follow for $N=2$. }

\centering{}\includegraphics[scale=1.42]{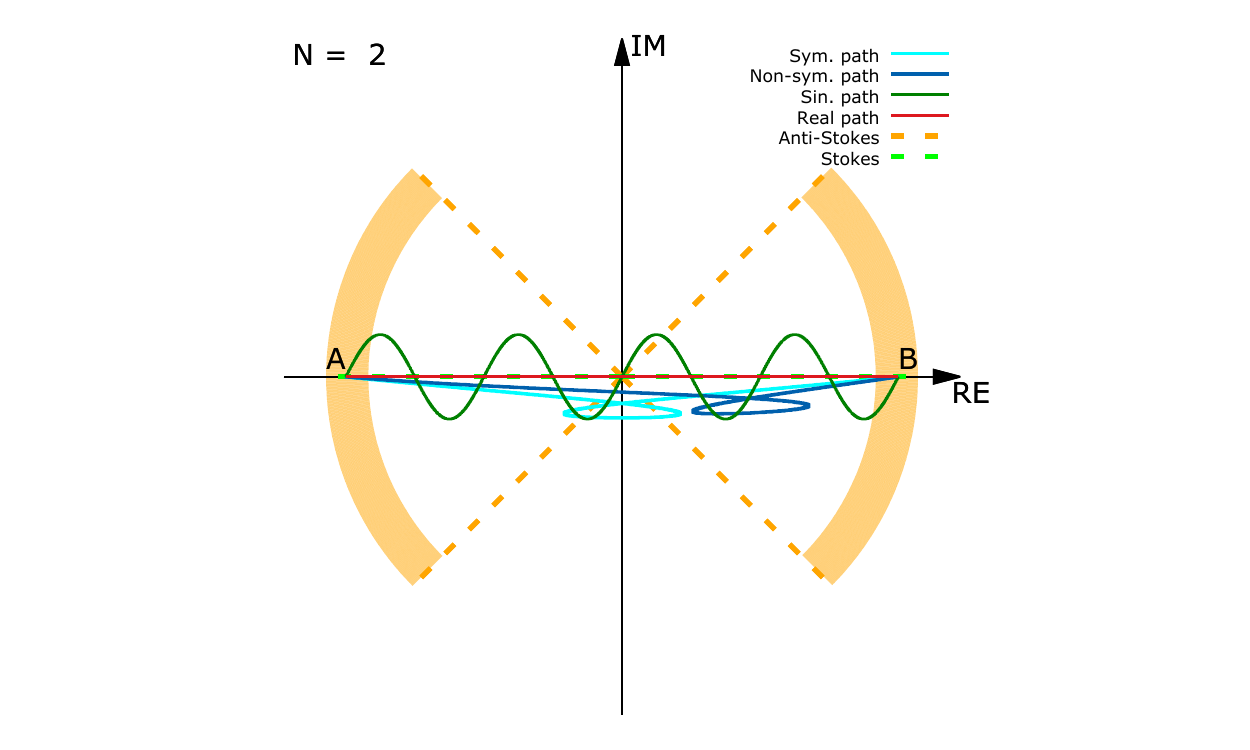}
\end{figure}

The first step to answer these questions is to parametrize various
paths and calculate the corresponding eigenvalues to see if they agree
each other or not. We start with the case when $N=2$ as a harmonic
oscillator, and use LMA and GLI to search for the eigenvalue by integrating
along the four different paths shown on Fig.\eqref{fig:indepdent_paths_N=00003D2},
where two symmetric points $A$ and $B$ are located on the real axis
and treated as two numerical infinities. These four different paths
start from the same boundary point $A$ and end to the same boundary
point $B$. One of the four paths is along the real axis, two complex
paths (sym. path and non-sym. path) are defined by \prettyref{eq:curve_cross_itself},
and we add another complex path (sin. path) defined by a sinusoidal
function. 

\begin{figure}[H]
\caption{\label{fig:curve_croses_itself}Parametric curve crosses itself.}

\centering{}\includegraphics[scale=0.55]{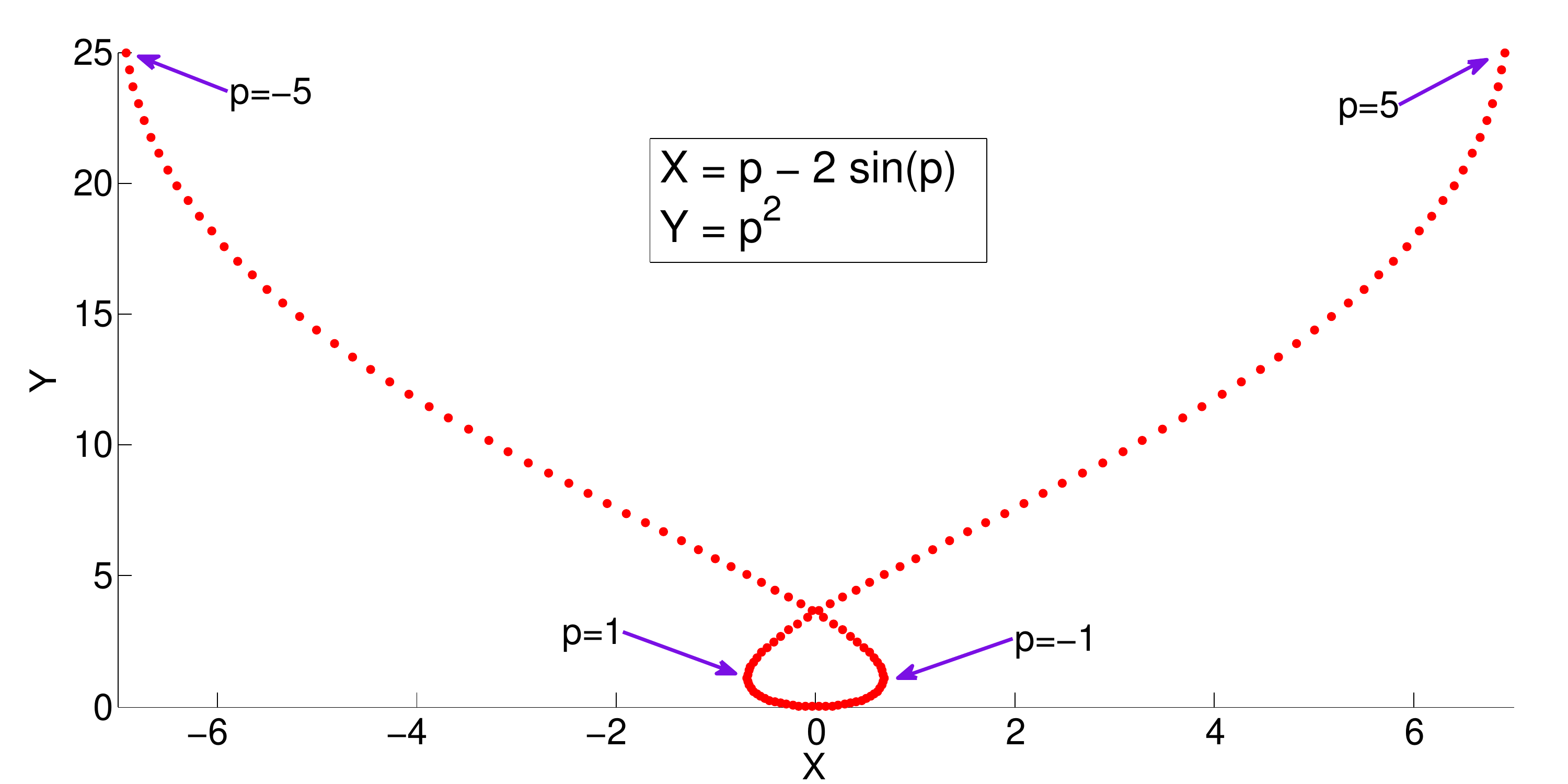}
\end{figure}

It's a bit of a challenge to parametrize the non-sym. path on Fig.\eqref{fig:indepdent_paths_N=00003D2}.
We start with the following parametric curve with real parameter $p$
shown on Fig.\eqref{fig:curve_croses_itself}

\begin{equation}
\begin{cases}
X=p-2\sin p\\
Y=p^{2}
\end{cases}\text{,}\label{eq:curve_cross_itself}
\end{equation}
which crosses itself once and is symmetric with respect to the $Y$-axis.
So this curve can be used to define the sym. path on Fig.\eqref{fig:indepdent_paths_N=00003D2}.
To get the non-symmetric version, we can just rotate the curve by
an given angle. 

The eigenvalues we found by following all four paths defined by Fig.\eqref{fig:indepdent_paths_N=00003D2}
are shown on the following Tab.\eqref{tab:different_path_N2}, where
the corresponding residues are obtained by calculating the complex
modulus of the numerical eigenfunctions at the right boundary point
$B$. Ideally, these values should be zero, so we set our tolerance
of residue to be $10^{-13}$. Through observation, the real parts
of eigenvalues are almost the same for all four paths, whereas the
imaginary parts are so small that they can be ignored. When the energy
level increases, both real and imaginary parts of eigenvalues $E$
start to shift from the analytic results, which are $E_{i}\in\left\{ 1,3,5,7\cdots\right\} $.
The higher energy level, the larger the shift. This is typical, since
we fix the locations of boundary points $A$ and $B$, and define
them as numerical infinities. When energy level becomes higher, the
pattern of corresponding eigenfunctions becomes more complicated -
they wiggle more before vanishing at the infinity so that the length
of non-vanishing parts becomes longer. Ideally, to minimize the shift,
we need to separate $A$ and $B$ even farther to accommodate longer
eigenfunction for higher energy level. For those, who are meticulous,
you may observe from Tab.\eqref{tab:different_path_N2} that following
the real path has relatively smaller imaginary parts of eigenvalues
than following complex paths. The reason is the same as what we have
just said. The non-vanishing parts of corresponding eigenfunctions
by following the complex paths are longer than by following the real
path, because the shapes of these complex paths are more complicated
than of the real path. In example from the next Sec.\prettyref{subsec:Six-distinct-paths},
we will numerically demonstrate that within the same pair of Stokes
wedges and independent from the shape of path, imaginary parts of
eigenvalues become smaller by separating $A$ and $B$ farther. 

\begin{table}[H]
\caption{\label{tab:different_path_N2}Eigenvalues $E$ from four distinct
paths for $N=2$. }

\centering{}%
\begin{tabular}{|c|c|c|c|}
\hline 
 & $\operatorname{Re}\left(E\right)$ & $\operatorname{Im}\left(E\right)$ & Residue\tabularnewline
\hline 
\hline 
Sym. path & 1.000000000000000 & -0.1E-17 & 0.6E-14\tabularnewline
\hline 
Non-sym. path & 1.000000000000000 & -0.9E-16 & 0.6E-15\tabularnewline
\hline 
Sin. path & 1.000000000000000 & -0.1E-19 & 0.6E-13\tabularnewline
\hline 
Real path & 1.000000000000000  & -0.6E-24 & 0.4E-13\tabularnewline
\hline 
\hline Sym. path & 3.000000000000000 & -0.3E-17 &  0.7E-14 \tabularnewline
\hline 
Non-sym. path & 3.000000000000002 & -0.1E-14 & 0.6E-13\tabularnewline
\hline 
Sin. path & 3.000000000000000  & -0.3E-16 &  0.2E-13 \tabularnewline
\hline 
Real path & 3.000000000000000 & -0.9E-25 &  0.4E-16 \tabularnewline
\hline 
\hline Sym. path & 5.000000000000013 & 0.2E-15 &  0.2E-13\tabularnewline
\hline 
Non-sym. path & 5.000000000000021 & -0.5E-14 &  0.7E-15\tabularnewline
\hline 
Sin. path & 5.000000000000013  & -0.1E-14  & 0.2E-13 \tabularnewline
\hline 
Real Path & 5.000000000000013 & -0.3E-21 &  0.4E-14 \tabularnewline
\hline 
\hline Sym. path & 7.000000000000336 & 0.7E-14 &  0.8E-14\tabularnewline
\hline 
Non-sym. path & 7.000000000000333 & -0.2E-13 &  0.9E-14 \tabularnewline
\hline 
Sin. path & 7.000000000000346  & -0.3E-13  & 0.5E-13\tabularnewline
\hline 
Real Path & 7.000000000000349 & -0.1E-21 &  0.5E-16 \tabularnewline
\hline 
\hline Sym. path & 9.000000000006479 & 0.1E-12 &  0.9E-13\tabularnewline
\hline 
Non-sym. path & 9.000000000005852 & -0.5E-13 &  0.6E-15 \tabularnewline
\hline 
Sin. path & 9.000000000007047 & -0.4E-13  & 0.3E-14 \tabularnewline
\hline 
Real path & 9.000000000006735 & -0.5E-18 &  0.1E-13 \tabularnewline
\hline 
\hline Sym. path & 11.00000000009739 & 0.2E-11 &  0.6E-13\tabularnewline
\hline 
Non-sym. path & 11.00000000008690 & -0.9E-14 &  0.1E-13\tabularnewline
\hline 
Sin. path & 11.00000000010035  & -0.8E-11  & 0.2E-13\tabularnewline
\hline 
Real path & 11.00000000010117  & -0.2E-18 & 0.4E-15\tabularnewline
\hline 
\hline Sym. path & 13.00000000118508 & 0.2E-10  & 0.6E-13\tabularnewline
\hline 
Non-sym. path & 13.00000000105809 & 0.1E-11 &  0.1E-13 \tabularnewline
\hline 
Sin. path & 13.00000000122042 & -0.1E-09  & 0.3E-13\tabularnewline
\hline 
Real path & 13.00000000122972 & -0.1E-18 &  0.1E-16 \tabularnewline
\hline 
\end{tabular}
\end{table}

\begin{figure}[H]
\caption{\label{fig:indepdent_paths_N=00003D2-1-1}$\operatorname{Re}\left(x\right)$
versus the eigenfunction of the ground level along the real and sinusoidal
path for $N=2$ (Before using \prettyref{eq:normalized_wave_forumula}
to normalize).}

\centering{}\includegraphics[scale=1.31]{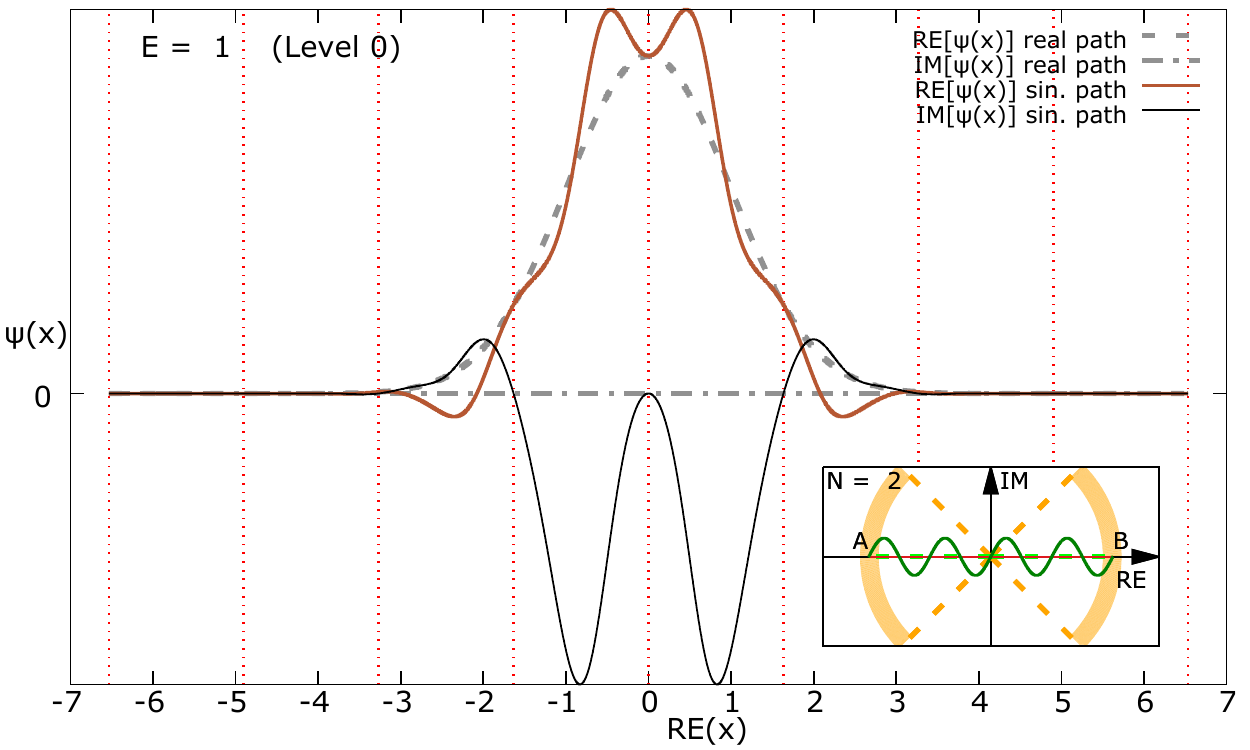}
\end{figure}

\begin{figure}[H]
\caption{\label{fig:indepdent_paths_N=00003D2-1-2}$\operatorname{Re}\left(x\right)$
versus the eigenfunction of the ground level along the real and sinusoidal
path for $N=2$ (after using \prettyref{eq:normalized_wave_forumula}
to normalize). }

\centering{}\includegraphics[scale=1.31]{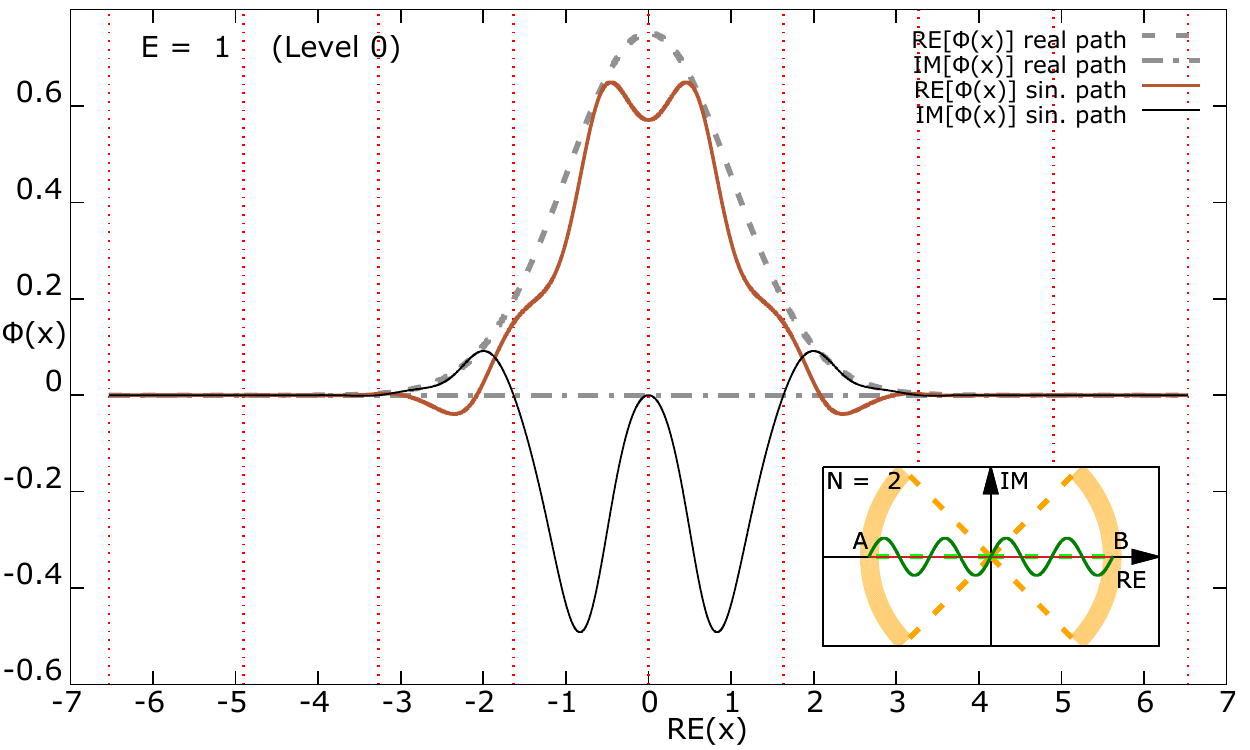}
\end{figure}

\begin{figure}[H]
\caption{\label{fig:indepdent_paths_N=00003D2-1-3}$\operatorname{Re}\left(x\right)$
or $p$ versus the eigenfunction of the ground level along three paths
(real, sin. and sym. path) for $N=2$. }

\centering{}\includegraphics[scale=1.31]{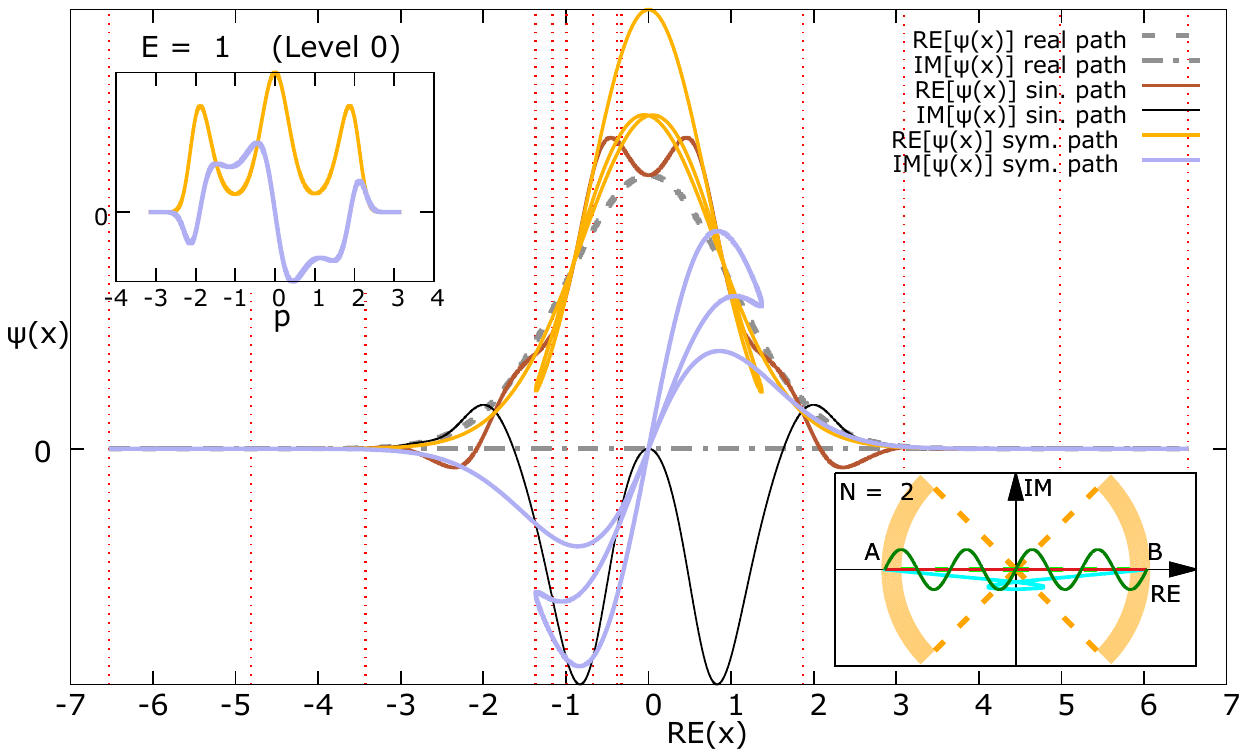}
\end{figure}

\begin{figure}[H]
\caption{\label{fig:indepdent_paths_N=00003D2-1-4}$\operatorname{Im}\left(x\right)$
versus the eigenfunction of the ground level along three paths (real,
sin. and sym. path) for $N=2$. }

\centering{}\includegraphics[scale=1.31]{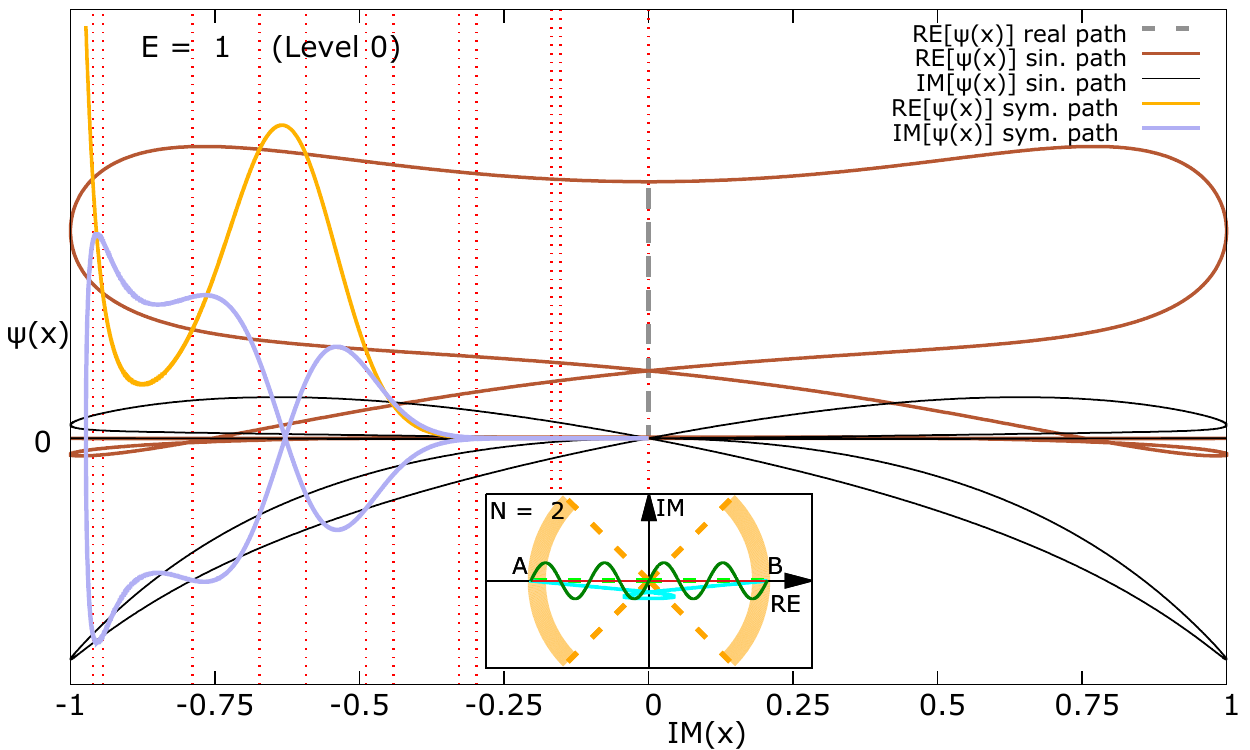}
\end{figure}

\begin{figure}[H]
\caption{\label{fig:indepdent_paths_N=00003D2-1-7}$\operatorname{Im}\left(x\right)$
versus the eigenfunction of the ground level along three paths (real,
sin. and non-sym. path) for $N=2$. }

\centering{}\includegraphics[scale=1.31]{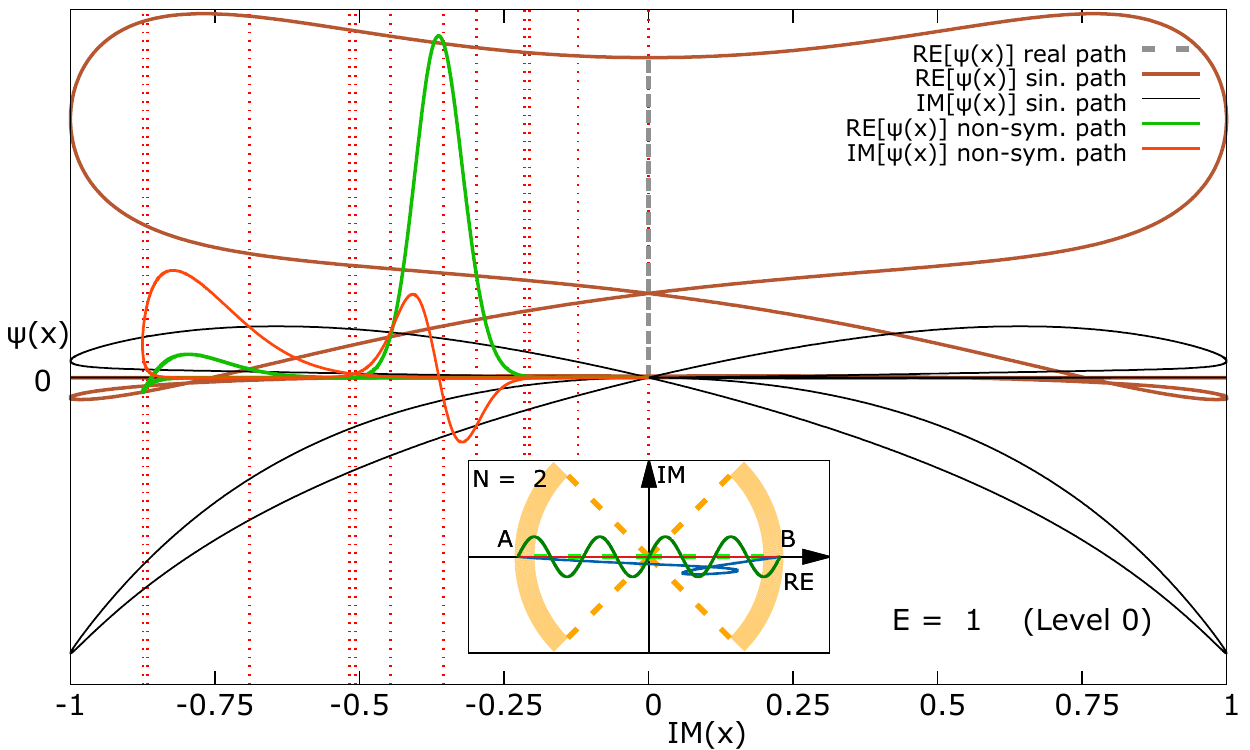}
\end{figure}

\begin{figure}[H]
\caption{\label{fig:indepdent_paths_N=00003D2-1-10}$\operatorname{Im}\left(x\right)$
versus the eigenfunction of the $4$th level along three paths (real,
sin. and non-sym. path) for $N=2$. }

\centering{}\includegraphics[scale=1.31]{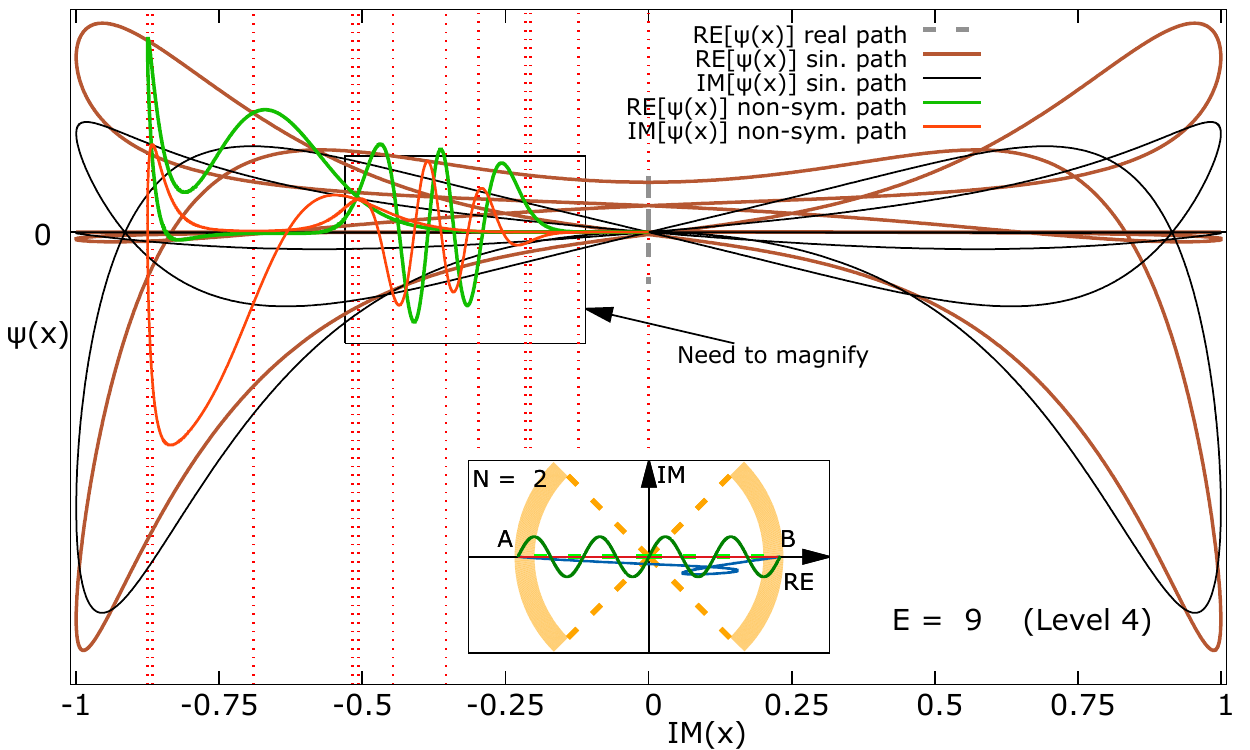}
\end{figure}

Now we plot the eigenfunctions associated with different paths. Fig.\eqref{fig:indepdent_paths_N=00003D2-1-3}
shows two mini-plots. One mini-plot is to show the shape of paths,
while the other shows the parameter $p$ defined by \prettyref{eq:curve_cross_itself}
versus the corresponding eigenfunction $\psi\left(p\right)$ by following
the sym. path. For $N=2$, since all imaginary parts of eigenfunctions
associated with the real path are zero, that is why we only see a
vertical segment (the grey dash line) contributed by $\operatorname{Re}\left[\psi\left(x\right)\right]$
along the real path on Fig.\eqref{fig:indepdent_paths_N=00003D2-1-4},
Fig.\eqref{fig:indepdent_paths_N=00003D2-1-7} and Fig.\eqref{fig:indepdent_paths_N=00003D2-1-10}.

\begin{figure}[H]
\caption{\label{fig:indepdent_paths_N=00003D2-1-5}$\operatorname{Re}\left(x\right)$
or $p$ versus the eigenfunction of the ground level along three paths
(real, sin. and non-sym. path) for $N=2$. }

\centering{}\includegraphics[scale=1.31]{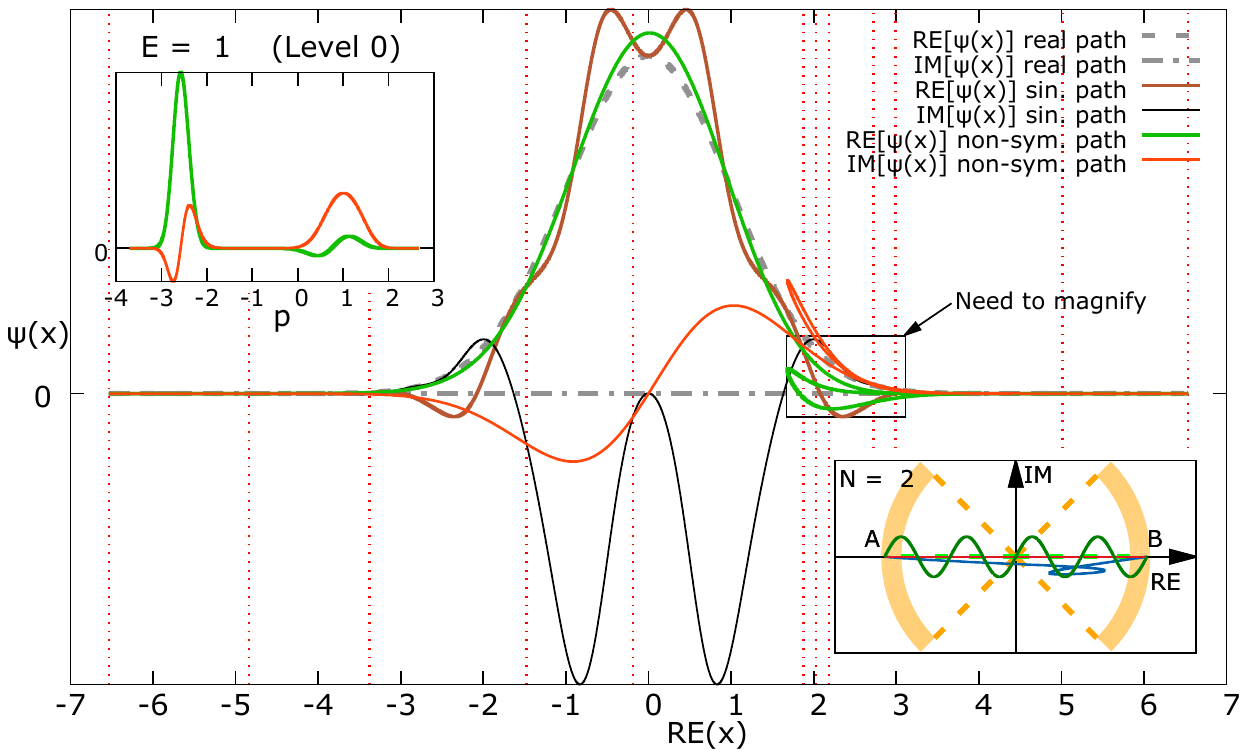}
\end{figure}

\begin{figure}[H]
\caption{\label{fig:indepdent_paths_N=00003D2-1-6}$\operatorname{Re}\left(x\right)$
versus the eigenfunction of the ground level along three paths (real,
sin. and non-sym. path) for $N=2$ (after magnifying). }

\centering{}\includegraphics[scale=1.31]{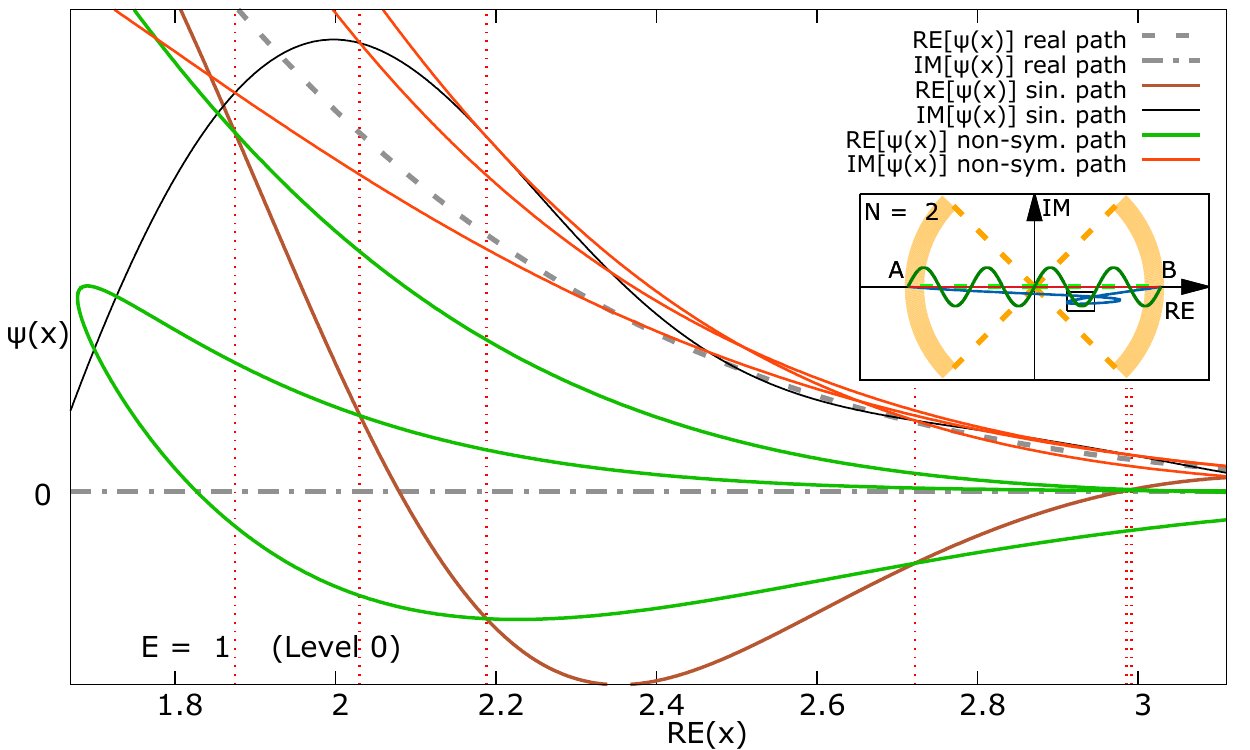}
\end{figure}

\begin{figure}[H]
\caption{\label{fig:indepdent_paths_N=00003D2-1-8}$\operatorname{Re}\left(x\right)$
or $p$ versus the eigenfunction of the $4$th level along three paths
(real, sin. and non-sym. path) for $N=2$. }

\centering{}\includegraphics[scale=1.29]{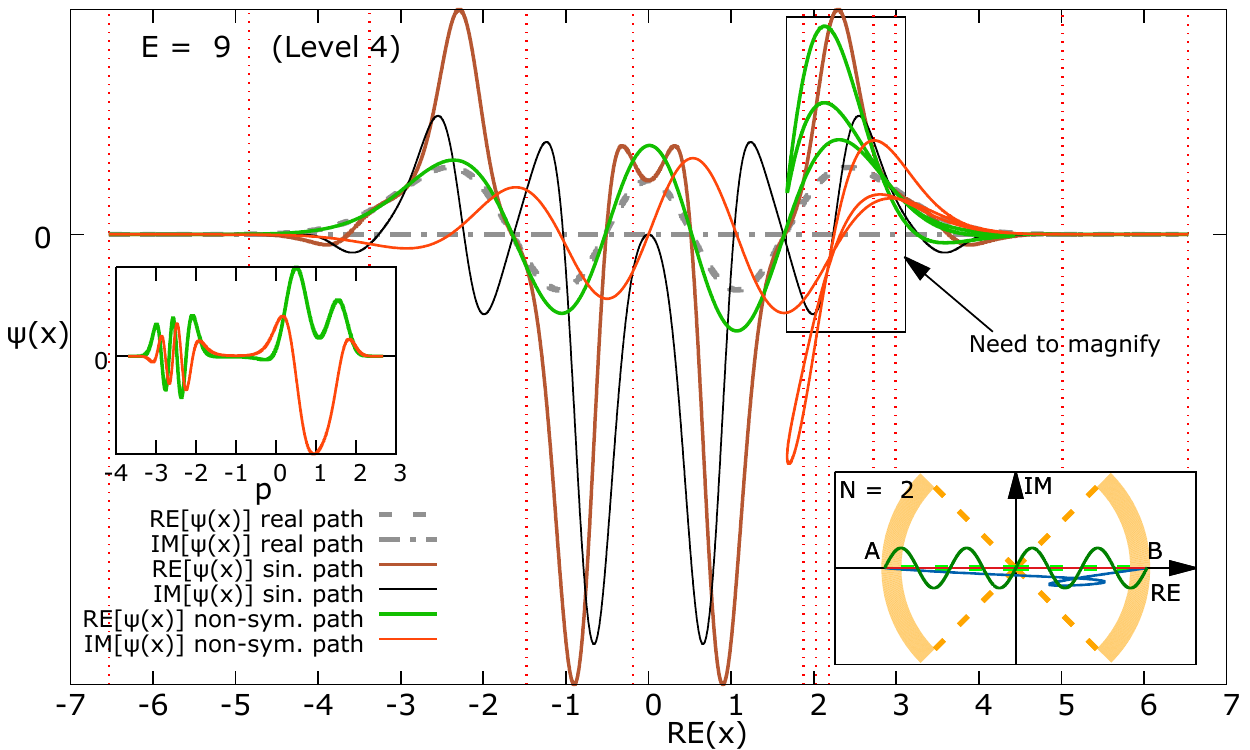}
\end{figure}

\begin{figure}[H]
\caption{\label{fig:indepdent_paths_N=00003D2-1-9}$\operatorname{Re}\left(x\right)$
versus the eigenfunction of the $4$th level along three paths (real,
sin. and non-sym. path) for $N=2$ (after magnifying). }

\centering{}\includegraphics[scale=1.29]{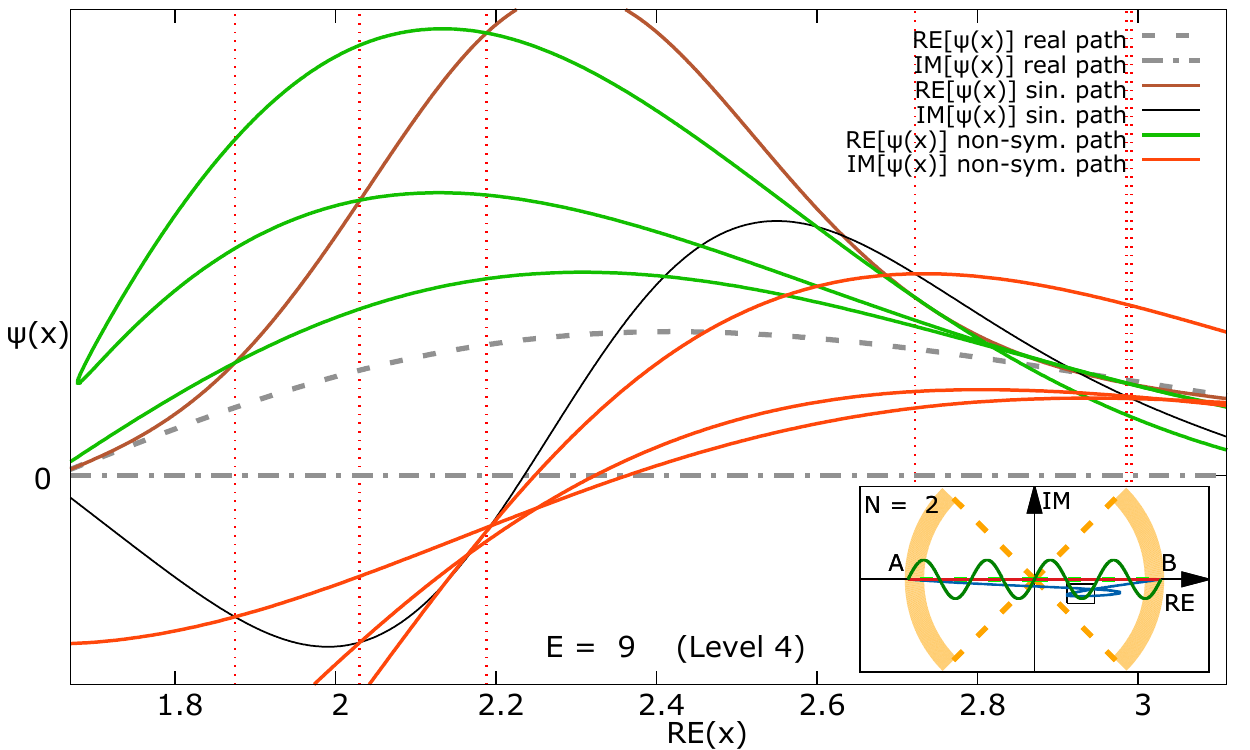}
\end{figure}

\begin{figure}[H]
\caption{\label{fig:indepdent_paths_N=00003D2-1-11}$\operatorname{Im}\left(x\right)$
versus the eigenfunction of the $4$th level along three paths (real,
sin. and non-sym. path) for $N=2$ (after magnifying). }

\centering{}\includegraphics[scale=1.31]{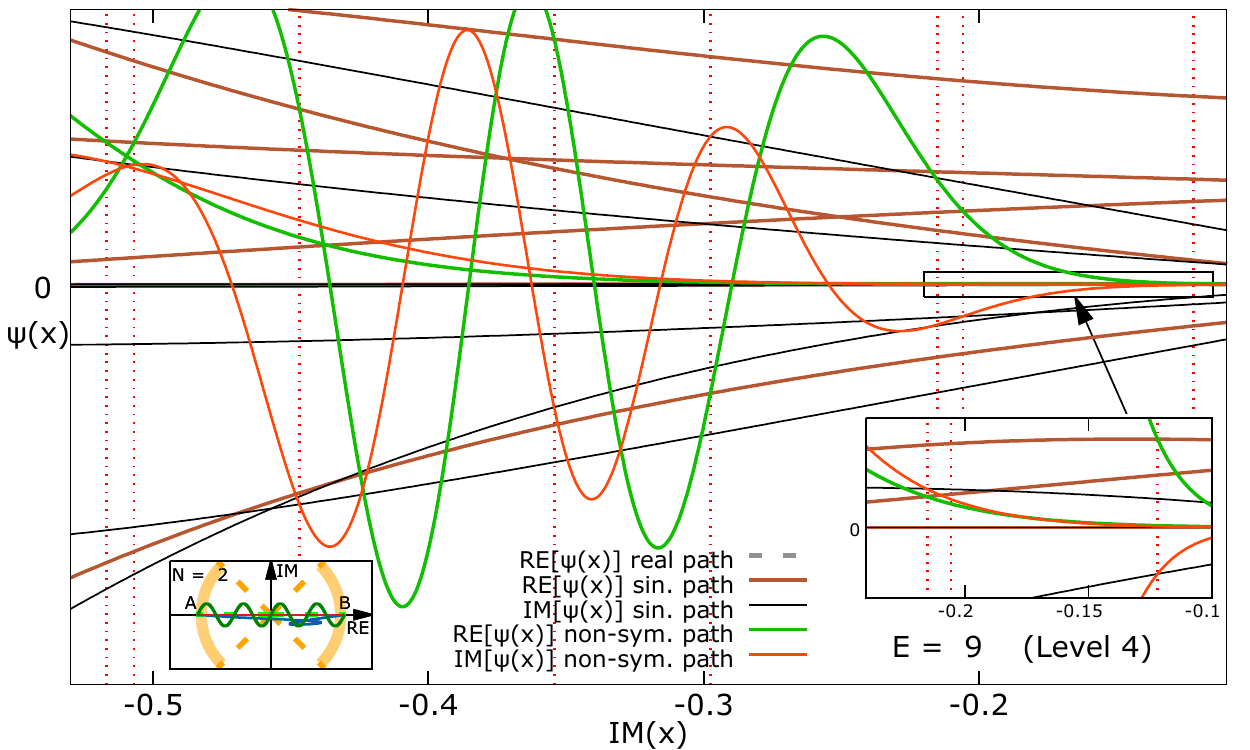}
\end{figure}

It seems that the different path has different eigenfunction even
though the eigenvalues are the same. This is an illusion! All corresponding
eigenfunctions are independent from the shape of paths as well. To
demonstrate this, we draw the vertical-red dotlines on all those figures
to indicate where two different paths intersect each other. At all
those intersection points, the corresponding two eigenfunctions cross
each other as well. We call these behaviors as ``crossing events''.
Suppose that two different paths intersect at one point $x_{*}$,
then a single crossing event is that the eigenfunction $\psi\left(x_{*}\right)$
from the one path and $\psi\left(x_{*}\right)$ from the other path
are crossed at $x_{*}$. For example, on Fig.\eqref{fig:indepdent_paths_N=00003D2-1-1},
$\operatorname{Re}\left[\psi\left(x\right)\right]$ from the real
path and $\operatorname{Re}\left[\psi\left(x\right)\right]$ from
the sinusoidal path are crossed at $9$ locations, and meanwhile $\operatorname{Im}\left[\psi\left(x\right)\right]$
from the real path and $\operatorname{Im}\left[\psi\left(x\right)\right]$
from the sinusoidal path are crossed at another $9$ locations. These
two sets of $9$ locations can be connected pair by pair by $9$ vertical-red
dotlines, whose horizontal coordinates are the real coordinates of
the $9$ intersection points between the two paths. On Fig.\eqref{fig:indepdent_paths_N=00003D2-1-4},
however, the horizontal coordinates are changed to be the imaginary
coordinates of the corresponding intersection points. By observation
on all figures from Fig.\eqref{fig:indepdent_paths_N=00003D2-1-3}
to Fig.\eqref{fig:indepdent_paths_N=00003D2-1-11}, we conclude that
the number of crossing events is equal to the number of vertical-red
dotlines, and further conclude that the eigenfunctions are independent
from the shape of paths so long as those paths all start from the
same boundary point $A$ and end at the same boundary point $B$.
In the later case (See Fig.\eqref{fig:indepdent_paths_N=00003D3}),
we will numerically demonstrate that the eigenfunction is not independent
from path if that path starts and ends on different boundary point.

The amplitude of eigenfunction depends on the shape of path. If a
path contains some points whose distances are far away from the origin,
then the amplitude of the corresponding eigenfunction must be large.
That is why the amplitude along the real axis is the smallest; whereas
the amplitude along the sym. path (on Fig.\eqref{fig:indepdent_paths_N=00003D2-1-3}
and Fig.\eqref{fig:indepdent_paths_N=00003D2-1-4}) is the largest. 

How about normalization? We're tempted to use standard normalization
from conventional quantum mechanics. In numerical approximation, the
normalized wavefunction $\phi_{n}\left(x\right)$ would be

\begin{equation}
\phi_{n}\left(x\right)=\frac{\psi_{n}\left(x\right)}{\sqrt{\int_{c}\psi_{n}^{*}\left(x\right)\psi_{n}\left(x\right)dx}}\approx\frac{\psi_{n}\left(p\right)}{\sqrt{\int_{p_{1}}^{p_{2}}\psi_{n}^{*}\left(p\right)\psi_{n}\left(p\right)dp}}\approx\frac{\psi_{n}\left(p\right)}{\sqrt{\underset{p}{\sum}\psi_{n}^{*}\left(p\right)\cdot\psi_{n}\left(p\right)\cdot dp}}\text{.}\label{eq:normalized_wave_forumula}
\end{equation}
where $p$ is the real parameter which parametrizes the path. By this
way, we find that $\phi_{n}\left(x\right)$ satisfies
\begin{equation}
\int_{c}\phi_{n}^{*}\left(x\right)\phi_{n}\left(x\right)dx\approx\int_{p_{1}}^{p_{2}}\phi_{n}^{*}\left(p\right)\phi_{n}\left(p\right)dp\approx\underset{p}{\sum}\phi_{n}^{*}\left(p\right)\cdot\phi_{n}\left(p\right)\cdot dp=1\text{ .}\label{eq:normalized_probability_formula}
\end{equation}
for all wave functions from different energy level and different path.
Hence, it is possible to ``normalize''\footnote{However, we are not able to use $\operatorname{Re}\left(x\right)$
to ``normalize'' $\psi\left(x\right)$ if $\operatorname{Re}\left(x\right)$
is not used to parametrize the contour path (For example, see \prettyref{eq:curve_cross_itself}).} all wave functions we have encountered so far! However, except of
the case by following the real path when $N=2$, the condition of
orthogonality $\int_{c}\phi_{m}^{*}\left(x\right)\phi_{n}\left(x\right)dx=0$
may not hold to be true in all other paths. For example, following
the sinusoidal path on Fig.\eqref{fig:indepdent_paths_N=00003D2-1-2},
we find that, numerically, 
\[
\int_{c}\phi_{0}^{*}\left(x\right)\phi_{1}\left(x\right)dx\approx0\qquad\int_{c}\phi_{6}^{*}\left(x\right)\phi_{3}\left(x\right)dx\approx0\qquad\int_{c}\phi_{0}^{*}\left(x\right)\phi_{5}\left(x\right)dx\approx0\text{,}
\]
but 
\[
\int_{c}\phi_{0}^{*}\left(x\right)\phi_{2}\left(x\right)dx=-0.92145+1.56493i\neq0\text{.}
\]

The ``official'' way of normalization introduced by Bender\cite{Make_sense}
is to use the recipe, which at first requires to find the $PT$-normalized
eigenfunction through 
\begin{equation}
\phi_{n}\left(x\right)=\exp\left(\nicefrac{i\theta_{n}}{2}\right)\psi_{n}\left(x\right)\text{,}\label{eq:PT-normalized_wavefunction}
\end{equation}
which satisfies $\phi_{n}^{*}\left(-x\right)=\phi_{n}\left(x\right)$.
Then we can verify 
\begin{equation}
\int_{c}\phi_{n}\left(x\right)\phi_{n}\left(x\right)dx=\left(-1\right)^{n}\text{.}\label{eq:PT_norm_value_chatper2}
\end{equation}
After that, we can use $CPT$-normalization defined by
\begin{equation}
\left\langle \phi_{m}\left(x\right),\phi_{n}\left(x\right)\right\rangle _{CPT}=\int_{c}\int_{c^{'}}\hat{C}\left(x,y\right)\phi_{m}\left(y\right)dy\phi_{n}\left(x\right)dx=\delta_{mn}\text{ ,}\label{eq:CPT_normalization}
\end{equation}
which in some case may require to find the charge operator $\hat{C}$
first. For the potential $-\left(ix\right)^{N}$, the most difficult
part is to find the phase angle $\theta_{n}$ from \prettyref{eq:PT-normalized_wavefunction}.
In this paper, we made no attempt to find $\theta_{n}$, and therefore
no attempt to normalize any eigenfunction we have encountered. 

Since \prettyref{eq:normalized_wave_forumula} is only a numerical
approximation of the normalization from conventional quantum mechanics,
Fig.\eqref{fig:indepdent_paths_N=00003D2-1-2} indicates that due
to numerical error or using the conventional/wrong method to normalize,
the two eigenfunctions no longer cross each other precisely at those
intersection points between the two paths. This means that the crossing
events no longer happen. We believe that such conundrum will still
exist even if we undertake the procedure of the $PT$-normalization
initiated at \prettyref{eq:PT-normalized_wavefunction}, because eventually
\prettyref{eq:PT-normalized_wavefunction} and \prettyref{eq:CPT_normalization}
require us to use numerical approximation again. This adds another
reason why we don't normalize eigenfunctions.

For higher energy state, we plot the $4$th level on the following
figures where we observe that the number of crossing events is equal
to the number of intersection points between two different paths as
well.

\subsubsection{When $N=3$ and $N=2.9$\label{subsec:Six-distinct-paths}}

\begin{figure}[H]
\caption{\label{fig:indepdent_paths_N=00003D3}Six distinct contour paths we
follow for $N=3$. }

\centering{}\includegraphics[scale=1.42]{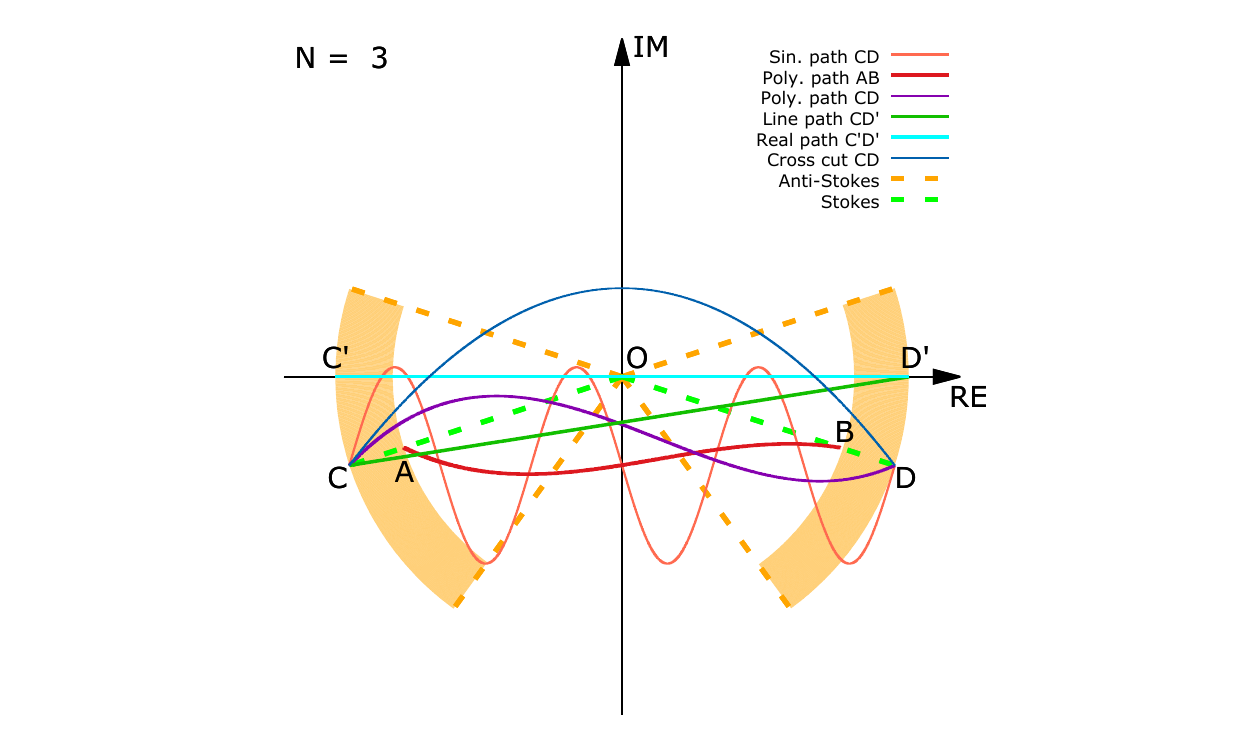}
\end{figure}

First, we draw three pairs of boundary points on Fig.\eqref{fig:indepdent_paths_N=00003D3}.
All these pairs $AB$, $CD$, and $C^{\prime}D^{\prime}$ are symmetric
with respect to the imaginary axis of $x$. If label the origin as
$O$, then $OA=OB$ and $OC=OD=OC^{\prime}=OD^{\prime}$. For further
test, we set six different paths for the case of $N=3$. Poly. path
$AB$ and poly. path $CD$ are parametrized by two different polynomials,
and both of paths are non-symmetric with respect to the imaginary
axis, and start on the left Stokes line and end on the right Stokes
line. In comparison, we add another four different paths on Fig.\eqref{fig:indepdent_paths_N=00003D3},
one of which is our old friend the sinusoidal path $CD$, the other
one from $C$ to $D$ crosses the positive-imaginary axis, and the
rest two are straight lines. One straight line path (real path $C^{\prime}D^{\prime}$)
is along the real axis from $C^{\prime}$ to $D^{\prime}$. The other
straight line path (line path $CD^{\prime}$) is slant, non-symmetric
and connects $C$ to $D^{\prime}$. 

\begin{table}[H]
\caption{\label{tab:different_path_N3}Eigenvalues of the $0$th and $1$st
level from six distinct paths for $N=3$. }

\centering{}%
\begin{tabular}{|c|c|c|c|}
\hline 
 & $\operatorname{Re}\left(E\right)$ & $\operatorname{Im}\left(E\right)$ & Residue\tabularnewline
\hline 
\hline 
Ploy. AB & 1.156267071989019  & 0.8E-14  & 0.5E-15\tabularnewline
\hline 
Poly. CD & 1.156267071988113 & 0.2E-23 & 0.3E-13 \tabularnewline
\hline 
Sin. CD & 1.156267071988113  & -0.1E-18  & 0.1E-13\tabularnewline
\hline 
Real path & 1.156267071988114 & -0.8E-17 & 0.4E-14\tabularnewline
\hline 
Line CD' & Unknown & Unknown & Unknown\tabularnewline
\hline 
Cross cut CD & 1.156267071988113  & 0.1E-22  & 0.1E-13 \tabularnewline
\hline 
\hline Ploy. AB & 4.109228752783768 & 0.2E-12 &  0.5E-15\tabularnewline
\hline 
Ploy. CD & 4.109228752809652 & 0.2E-21 &  0.8E-14 \tabularnewline
\hline 
Sin. CD & 4.109228752809652 & 0.1E-17  &  0.9E-14\tabularnewline
\hline 
Real path & 4.109228752809730 & -0.1E-15 &  0.1E-14\tabularnewline
\hline 
Line CD' & Unknown & Unknown & Unknown\tabularnewline
\hline 
Cross cut CD & 4.109228752809652 & -0.1E-22  & 0.2E-14\tabularnewline
\hline 
\end{tabular}
\end{table}

As shown on Tab.\eqref{tab:different_path_N3}, when we separate $AB$
even farther to $CD$, all imaginary parts of eigenvalues $E$ along
the poly. path become smaller. These demonstrate the claim we made
in the previous example. 

Since no real eigenvalue associated with the straight line path $CD^{\prime}$
is found, we then conclude that the two infinities $\infty_{left}$
and $\infty_{right}$ or two boundary points have to be symmetric
with respect to the imaginary axis to have real eigenvalue. 

\begin{figure}[H]
\caption{\label{fig:indepdent_paths_N=00003D3-5}$\mbox{Re}\left(x\right)$
versus the eigenfunction of the ground level along two paths (poly.
CD and sin. CD) for $N=3$. The number of crossing events is equal
to the number of intersection points between the two paths. }

\centering{}\includegraphics[scale=1.32]{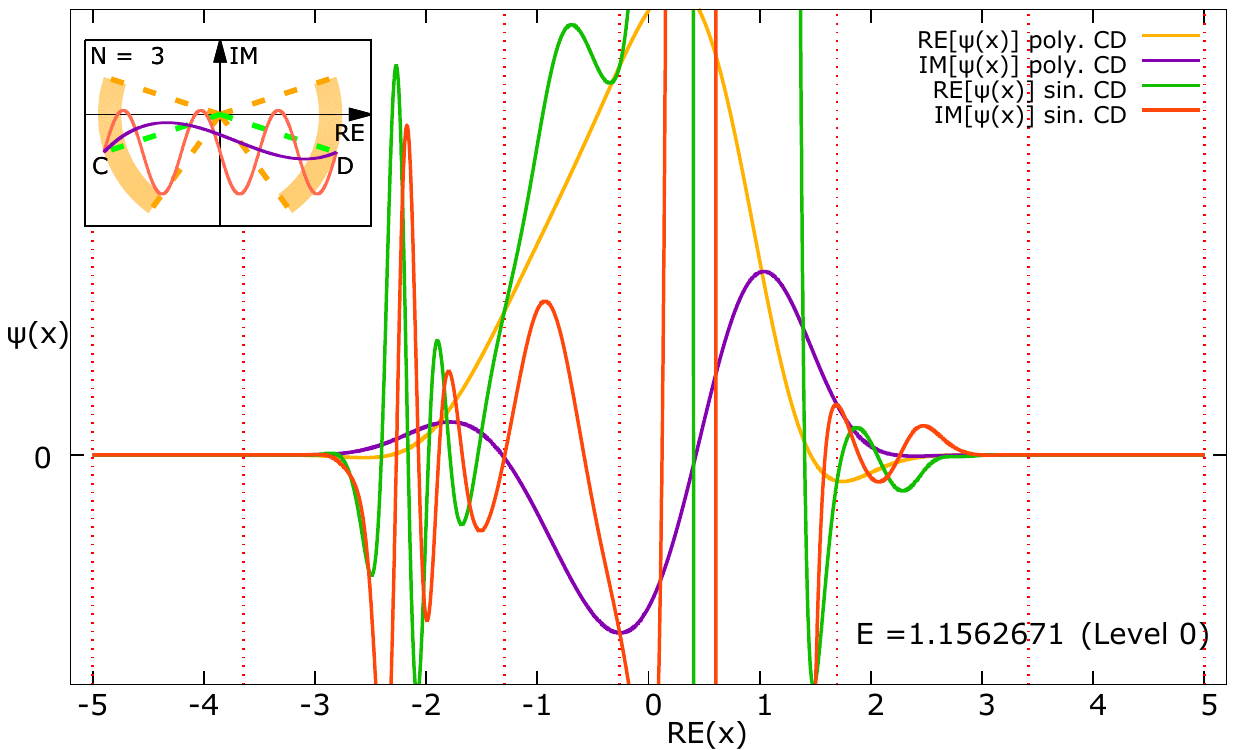}
\end{figure}

\begin{figure}[H]
\caption{\label{fig:indepdent_paths_N=00003D3-6}$\mbox{Re}\left(x\right)$
versus the eigenfunction of the ground level along two paths (poly.
AB and sin. CD) for $N=3$. No crossing event happens.}

\centering{}\includegraphics[scale=1.32]{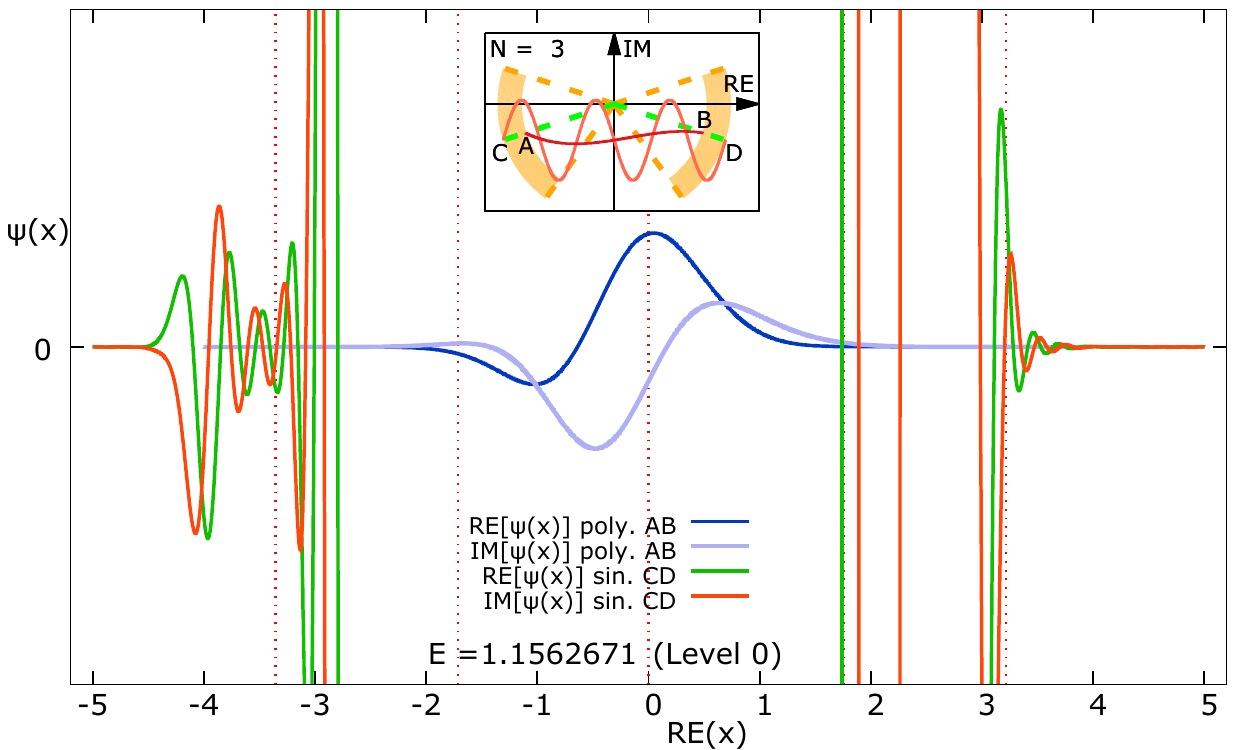}
\end{figure}

\begin{figure}[H]
\caption{\label{fig:indepdent_paths_N=00003D3-7}$\mbox{Re}\left(x\right)$
versus the eigenfunction of the ground level along two paths (sin.
CD and real C'D') for $N=3$. No crossing event happens.}

\centering{}\includegraphics[scale=1.32]{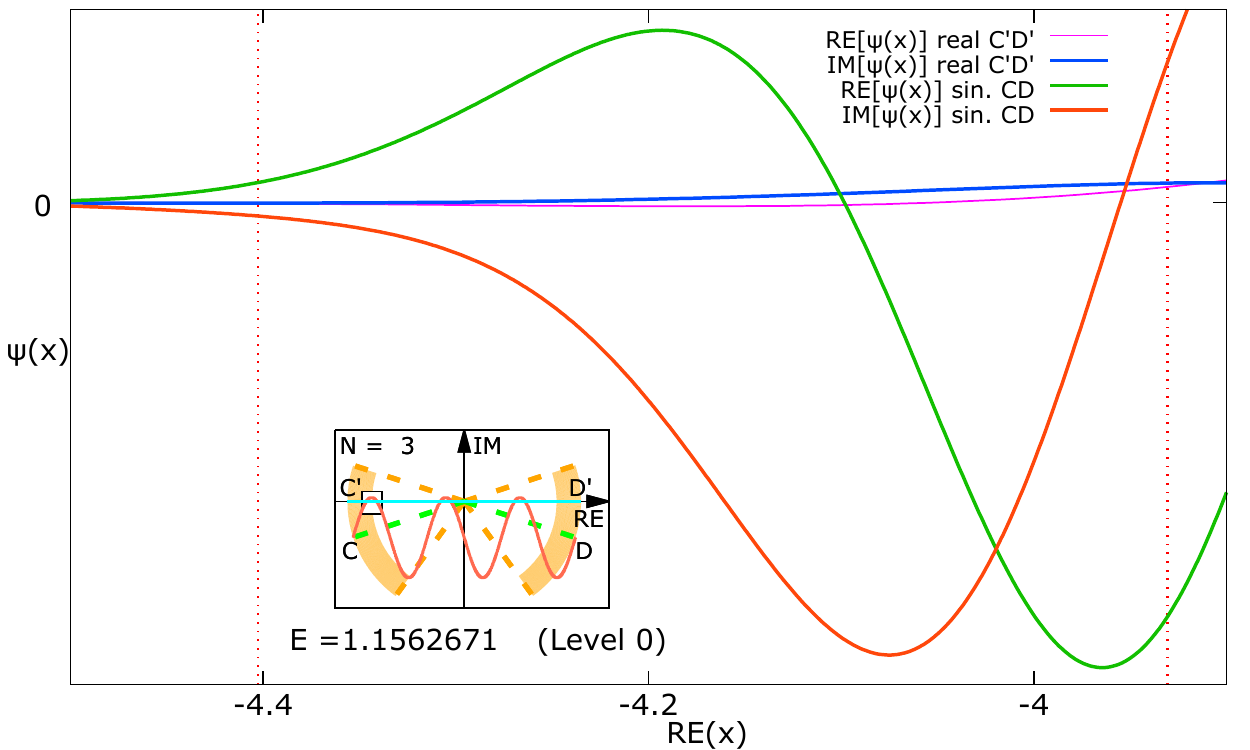}
\end{figure}

On Fig.\eqref{fig:indepdent_paths_N=00003D3-5}, Fig.\eqref{fig:indepdent_paths_N=00003D3-6}
and Fig.\eqref{fig:indepdent_paths_N=00003D3-7}, our purpose is not
to plot the entire eigenfunctions, but only to show whether the crossing
events occur or not. On Fig.\eqref{fig:indepdent_paths_N=00003D3-5},
we observe that the number of crossing events is equal to the number
of intersection points between the two paths. This is not only true
for integer $N$ but also for fractional $N$ (e.g. $N=2.9$). However,
if any two paths start and end at different boundary points within
a pair of Stokes wedges, then the crossing event will not happen -
for example, on Fig.\eqref{fig:indepdent_paths_N=00003D3-6} one path
goes from $C$ to $D$ while the other goes from $A$ to $B$. Since
$A$ and $B$ are closer to the origin, the amplitude of the wave
function is smaller so that no crossing event happens. Another example
is shown on Fig.\eqref{fig:indepdent_paths_N=00003D3-7}, where one
path goes from $C$ to $D$ while the other goes from $C^{\prime}$
to $D^{\prime}$, and no crossing event happens even though $OC=OD=OC^{\prime}=OD^{\prime}$.

\begin{figure}[H]
\caption{\label{fig:indepdent_paths_N=00003D29}Six distinct contour paths
we follow for $N=2.9$. }

\centering{}\includegraphics[scale=1.42]{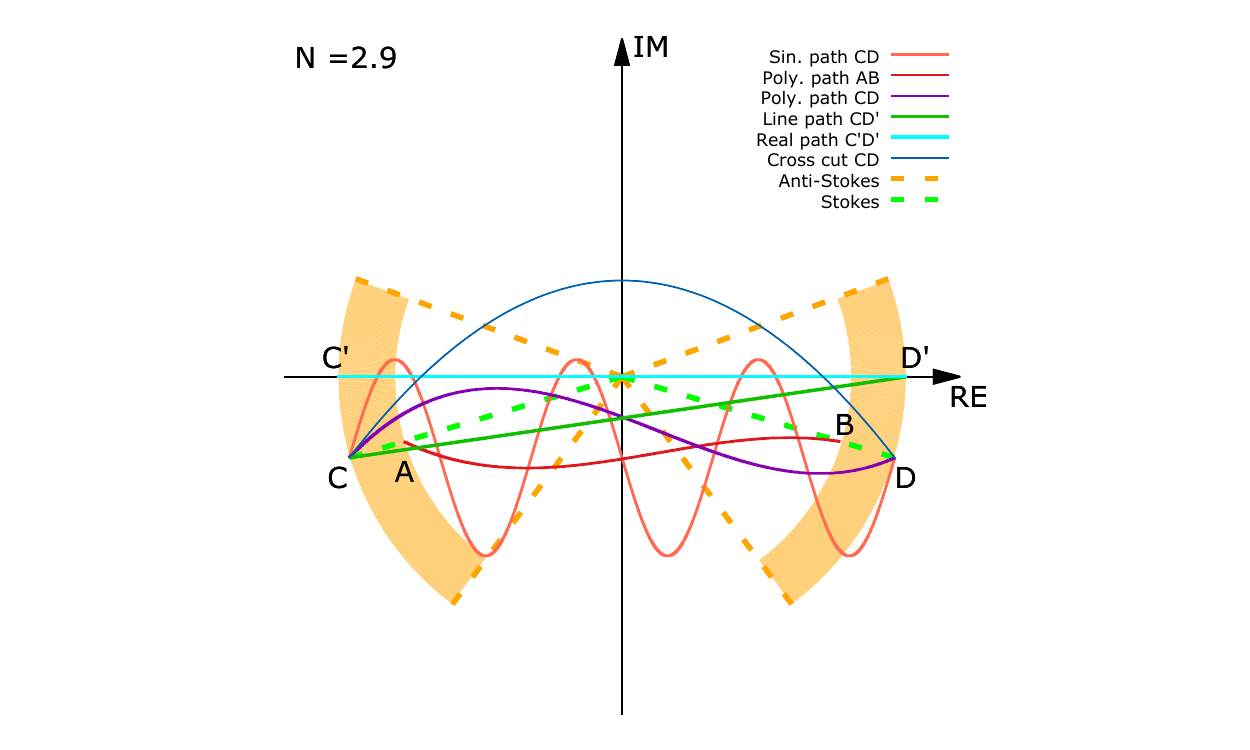}
\end{figure}

On Tab.\eqref{tab:different_path_N3}, it is a little surprise to
see that the path (cross cut $CD$) yields the same eigenvalues as
those paths without crossing the cut. The crossing events also happen
in this case, where the two paths (cross cut $CD$ and sin. $CD$)
are involved. However, it is not ``safe'' to cross the cut if $N$
is not an integer. For example, in case when $N=2.9$, the locations
of Stokes lines and anti-Stokes lines on Fig.\eqref{fig:indepdent_paths_N=00003D29}
are slightly changed, so that we shift the boundary points $A$, $B$,
$C$, $D$ accordingly and calculate eigenvalues again. Tab.\eqref{tab:different_path_N29}
shows that the eigenvalue associated with the path (cross cut $CD$)
is drastically changed even if $N$ is changed only by $0.1$. We
only find one real and negative eigenvalue. The rest eigenvalues may
be complex. Tab.\eqref{tab:different_path_N3} and Tab.\eqref{tab:different_path_N29}
imply that the path which crosses the cut on the positive-imaginary
axis must give the same eigenvalue as those paths without crossing
it, only if $N$ is an integer. 

\begin{table}[H]
\caption{\label{tab:different_path_N29}Eigenvalues of the $0$th and $1$st
level from six distinct paths for $N=2.9$. }

\centering{}%
\begin{tabular}{|c|c|c|c|}
\hline 
 & $\operatorname{Re}\left(E\right)$ & $\operatorname{Im}\left(E\right)$ & Residue\tabularnewline
\hline 
\hline 
Ploy. AB & 1.131396959784506  & 0.5E-13  & 0.8E-14\tabularnewline
\hline 
Poly. CD & 1.131396959777214 & 0.8E-22 & 0.9E-14 \tabularnewline
\hline 
Sin. CD & 1.131396959777214  & -0.8E-20 & 0.4E-13\tabularnewline
\hline 
Real path & 1.131396959777217 & -0.9E-16  & 0.5E-14 \tabularnewline
\hline 
Line CD' & Unknown & Unknown & Unknown\tabularnewline
\hline 
Cross cut CD & -0.1948727126451554  & -0.4E-22  & 0.2E-13\tabularnewline
\hline 
\hline Ploy. AB & 3.958636971974068 & 0.1E-11 &  0.1E-14\tabularnewline
\hline 
Ploy. CD & 3.958636972135053 & 0.1E-19  &  0.3E-13\tabularnewline
\hline 
Sin. CD & 3.958636972135053 & 0.1E-18  &  0.1E-13 \tabularnewline
\hline 
Real path & 3.958636972135127 & 0.3E-15 & 0.2E-13\tabularnewline
\hline 
Line CD' & Unknown & Unknown & Unknown\tabularnewline
\hline 
Cross cut CD & Unknown & Unknown & Unknown\tabularnewline
\hline 
\end{tabular}
\end{table}

\subsubsection{\label{subsec:Summary}Summary}
\begin{enumerate}
\item For the potential $-\left(ix\right)^{N}$ with $N>1$, one necessary
condition to have real-positive eigenvalue is that the two boundary
points for any path must be symmetric with respect to the imaginary
axis of $x$. 
\item If none of paths crosses the cut, and suppose that one path has boundary
points $A$ and $B$ symmetric with respect to the imaginary axis,
whereas the other path has boundary points $C$ and $D$ symmetric
with respect to the imaginary axis, and $A$, $B$, $C$, $D$ all
lie within the same pair of Stokes wedges, then eigenvalues for these
two paths must be the same even if $A\neq C$ and $B\neq D$. However,
their eigenfunctions may be different. 
\item Suppose that two paths have the same boundary points $A$ and $B$
symmetric with respect to the imaginary axis, one path crosses the
cut on the positive-imaginary axis and the other does not, and $A$,
$B$ lie within a pair of Stokes wedges, then their eigenvalues and
eigenfunctions must be all independent from the shape of path if $N$
is an integer; and dependent if $N$ is an non-integer.
\item Suppose that two paths have the same boundary points $A$ and $B$
symmetric with respect to the imaginary axis, none of the paths crosses
the cut on the positive-imaginary axis, and $A$, $B$ lie within
a pair of Stokes wedges, then their eigenvalues and eigenfunctions
must be all independent from the shape of path.
\end{enumerate}

\section{\label{sec:Multiple_families}Multiple families of real energy spectrum}

\subsection{Comparison between two pairs of $PT$-symmetric wedges when $N=5$}

\begin{figure}[H]
\caption{\label{fig:indepdent_paths_N=00003D5}Two distinct paths we follow
for $N=5$. }

\centering{}\includegraphics[scale=1.42]{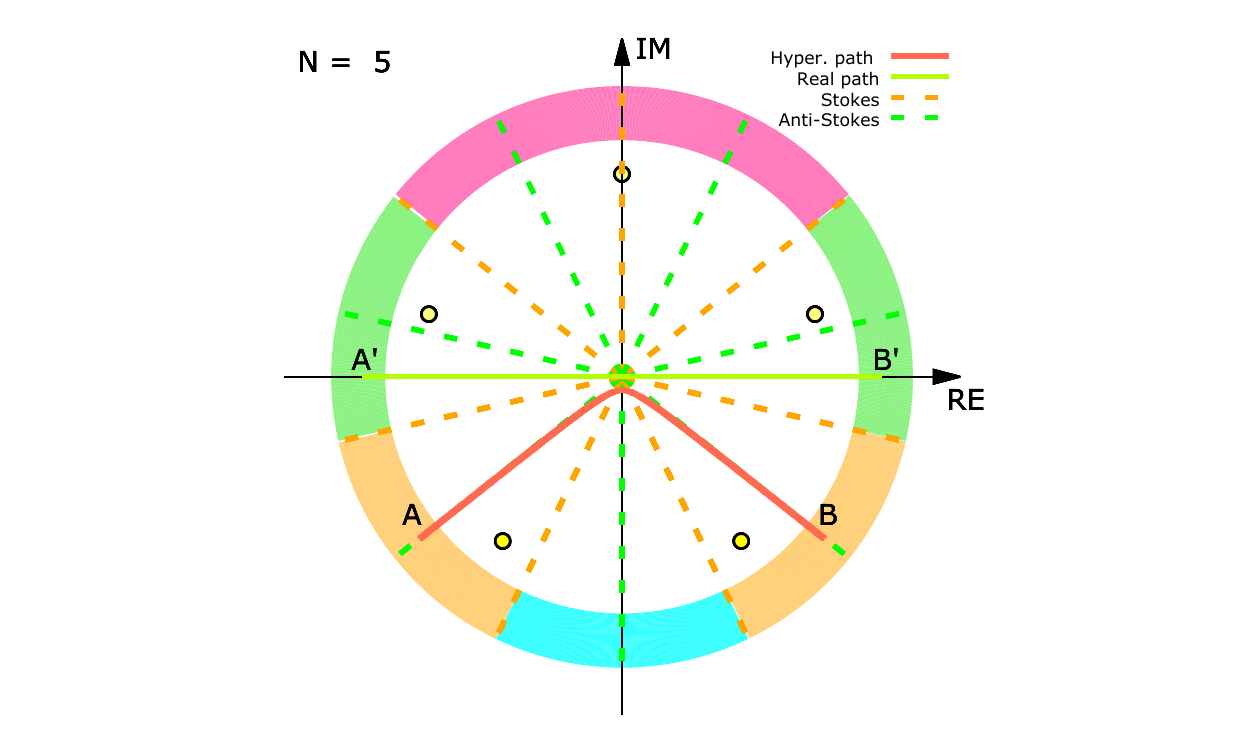}
\end{figure}

\begin{table}[H]
\caption{\label{tab:different_path_N5}Eigenvalues $E$ from the two distinct
paths for $N=5$. }

\centering{}%
\begin{tabular}{|c|c|c|c|c|}
\hline 
 & $\operatorname{Re}\left(E\right)$ & $\operatorname{Im}\left(E\right)$ & Residue & Ratio of $\operatorname{Re}\left(E\right)$\tabularnewline
\hline 
\hline 
Hyper. path & 1.908264578170778  & -0.1E-29  & 0.2E-15 & \multirow{2}{*}{1.638318217184208}\tabularnewline
\cline{1-4} 
Real path & 1.164770407943415 & -0.8E-21  & 0.4E-14 & \tabularnewline
\hline 
\hline Hyper. path & 8.587220836207222  & -0.1E-29  & 0.7E-15 & \multirow{2}{*}{1.967838030619607}\tabularnewline
\cline{1-4} 
Real path & 4.363784367712109 & -0.1E-19 & 0.4E-13 & \tabularnewline
\hline 
\hline Hyper. path & 17.71080901173115 & -0.4E-28  & 0.4E-14 & \multirow{2}{*}{1.977719568513977}\tabularnewline
\cline{1-4} 
Real Path & 8.955166998240672  & -0.2E-18 & 0.9E-13 & \tabularnewline
\hline 
\hline Hyper. path & 28.59510331173597 & 0.4E-27 & 0.5E-13 & \multirow{2}{*}{1.983325673682043}\tabularnewline
\cline{1-4} 
Real path & 14.41775483027413  & -0.1E-17  & 0.2E-13 & \tabularnewline
\hline 
\hline Hyper. path & 40.91889089052085 & -0.9E-26 & 0.1E-13 & \multirow{2}{*}{1.985376898653392}\tabularnewline
\cline{1-4} 
Real path & 20.61013751004891 & -0.1E-16  & 0.4E-14 & \tabularnewline
\hline 
\end{tabular}
\end{table}

A typical question is what if we follow a path whose boundary points
are outside of the chosen (orange) wedges. Can we find any real eigenvalue?
The answer is yes. As shown on Tab.\eqref{tab:different_path_N5},
we have one family of real spectrum by following the hyperbolic path
within the orange wedges and another family of real spectrum by the
real path within the green wedges on Fig.\eqref{fig:indepdent_paths_N=00003D5},
and the ratios between the two families approach to a constant as
the energy level increases. 

\subsection{\label{subsec:The-leading-order-WKB}The leading-order WKB approximation}

The constant ratio from Tab.\eqref{tab:different_path_N5} can be
predicted by conventional WKB approximation\cite{schmidt2013generation}.
For the first step to do this, it's tempted to find turning points
by equating $E$ and the potential from \prettyref{eq:Eigenvalue_eq},
so that 
\begin{equation}
E=-\left(ix\right)^{N}\Longrightarrow x=E^{\nicefrac{1}{N}}e^{i\left(\frac{2-N}{2N}+\frac{2j}{N}\right)\pi}\qquad\mbox{for }j=0,\pm1,\pm2\cdots\text{.}\label{eq:turing_points_complete_expression}
\end{equation}
However, this expression may overestimate the total number of turning
points on the principal branches. For example, \prettyref{eq:turing_points_complete_expression}
suggests $7$ turning points for $N=3.5$, however, only $4$ turning
points on the principal branches.

Instead of the expression from \prettyref{eq:turing_points_complete_expression},
let's write the turning points $\left\{ x_{i}\right\} $ from a principal
branch in terms of $x_{\pm}$ and
\begin{equation}
x_{-}=E^{\nicefrac{1}{N}}e^{i\beta}\text{,}\label{eq:turing_point_gamma}
\end{equation}
\begin{equation}
x_{+}=E^{\nicefrac{1}{N}}e^{i\gamma}\text{,}\label{eq:turing_point_beta}
\end{equation}
where $\beta\neq\gamma$ due to different values from $j$. By the
leading-order WKB approximation on the complex plane, 
\begin{align}
\intop_{x_{-}}^{x_{+}}dx\sqrt{E-V\left(x\right)} & =\left(n+\frac{1}{2}\right)\pi\text{,}\label{eq:leading_order_WKB}
\end{align}
we finally obtain\cite{Real_Spectra,schmidt2013generation} 
\begin{equation}
E=\left[\frac{2\left(n+\frac{1}{2}\right)\sqrt{\pi}\Gamma\left(\frac{3}{2}+\frac{1}{N}\right)}{\left(e^{i\gamma}-e^{i\beta}\right)\Gamma\left(1+\frac{1}{N}\right)}\right]^{\frac{2N}{N+2}}\text{,}\label{eq:leading_order_mid3}
\end{equation}
where $E$ can be real if 
\begin{equation}
\operatorname{Im}\left(e^{i\gamma}-e^{i\beta}\right)=0\text{,}\label{eq:Condition_for_real_eigen1}
\end{equation}
which implies that 
\begin{equation}
\beta=\pi-\gamma\text{.}\label{eq:PT_symmtric_angle}
\end{equation}
Therefore, the turning points $x_{-}$ and $x_{+}$ must be symmetrical
with respect to the imaginary axis to have real eigenvalue $E$. By
\prettyref{eq:PT_symmtric_angle}, we obtain the leading-order approximation
for the eigenvalue $E_{n}$ 
\begin{equation}
E_{n}\sim\left[\frac{\left(n+\frac{1}{2}\right)\sqrt{\pi}\Gamma\left(\frac{3}{2}+\frac{1}{N}\right)}{\cos\gamma\Gamma\left(1+\frac{1}{N}\right)}\right]^{\frac{2N}{N+2}}\qquad\mbox{for }n\rightarrow\infty\text{.}\label{eq:leading_order_eigenvalue_result_Fake}
\end{equation}
Suppose that we have two families of real spectra associated with
two pairs of Stokes wedges, then these two families must be also associated
with two pairs of turning points. Assume that one pair of turning
points is associated with $\gamma_{1}$ and the other pair with $\gamma_{2}$,
then by \prettyref{eq:leading_order_eigenvalue_result_Fake} we have\cite{schmidt2013generation}
\begin{equation}
\frac{E_{n}\left(\gamma_{2}\right)}{E_{n}\left(\gamma_{1}\right)}\sim\left[\frac{\cos\left(\gamma_{1}\right)}{\cos\left(\gamma_{2}\right)}\right]^{\frac{2N}{N+2}}\qquad\mbox{for }n\rightarrow\infty\text{.}\label{eq:leading_order_ratio}
\end{equation}
The values of real spectra from the two families must approach to
the constant ratio according to \eqref{eq:leading_order_ratio}. For
example, when $N=5$,
\begin{equation}
\left[\frac{\cos\left(\gamma_{1}\right)}{\cos\left(\gamma_{2}\right)}\right]^{\frac{2N}{N+2}}=\left[\frac{\cos\left(\frac{1}{10}\pi\right)}{\cos\left(-\frac{3}{10}\pi\right)}\right]^{\frac{10}{7}}=1.988629015490531\text{ .}\label{eq:N=00003D5_WKB_ratio}
\end{equation}
Our numerical results from Tab.\eqref{tab:different_path_N5} agrees
with this WKB approximation as the energy level $n$ increases.

So the conclusion\cite{schmidt2013generation} is that there exists
more than one family of real spectra if $N$ is large enough, and,
as energy level increases, one family of real spectrum over the other
family maybe approaches to a constant ratio, which sometimes can be
predicted by the WKB approximation. For integer $N$, how many families
of real spectra there are depends on how many pairs of symmetric turning
points there are or how many pairs of symmetric but non-contacting
wedges there are. We will discuss more in Sec.\eqref{sec:Multiple_families}
about the families of real spectra.

\subsection{Energy spectra from the first four families versus $N$ }

\begin{figure}[H]
\caption{\label{fig:final_result_eigenvalue1}Energy spectrum of the $1$st
family from the pair of the orange wedges (The grey curves are the
WKB approximation).}

\centering{}\includegraphics[scale=1.37]{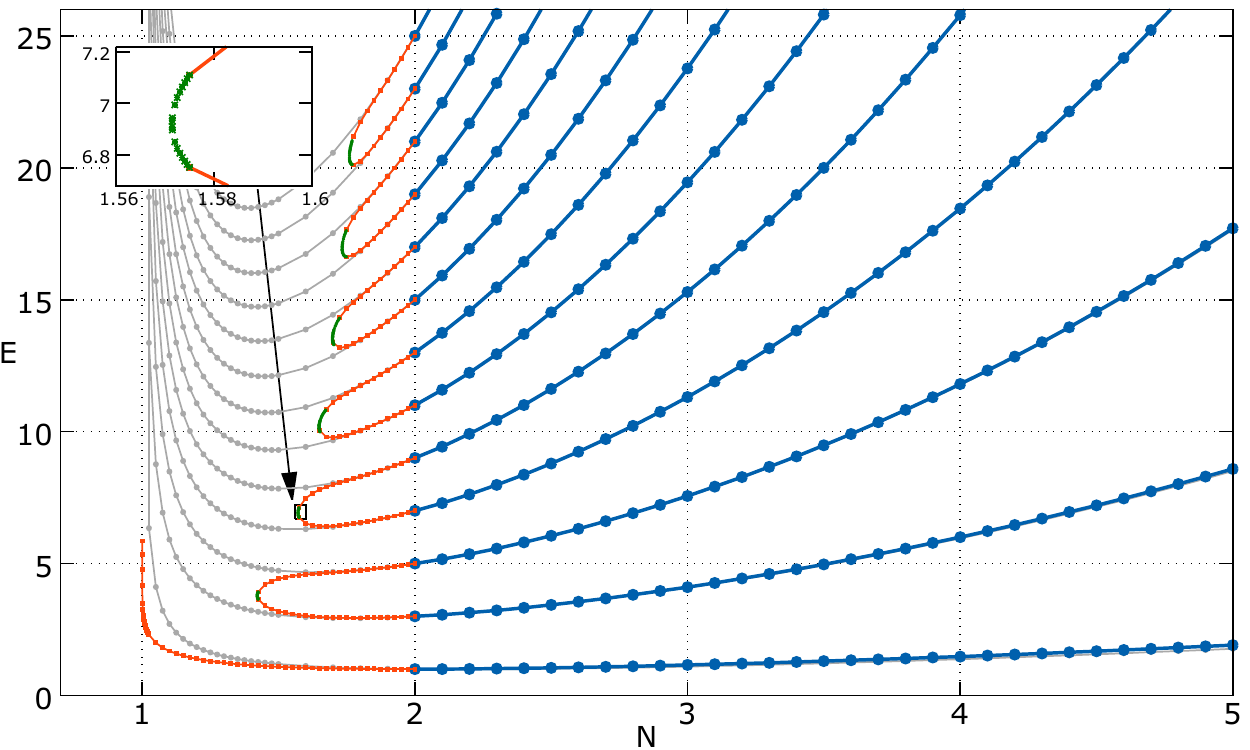}
\end{figure}

Whether the Stokes lines on the pair of the orange wedges (see Fig.\eqref{fig:chosen_wedges_multiplot})
move above or below the real axis depends on whether $N<2$ or $N>2$.
Based on this fact, in our code we set when $N\leq1.6$, we integrate
along the parametric path defined by \prettyref{eq:function_for_parametrix_Nless2};
when $1.6<N<3.0$, we follow the real axis; when $N\geq3.0$, the
hyperbolic path by \prettyref{eq:hypola_parametriz_eq} is followed.
Using LMA and GLI with the boundary condition applied within the orange
wedges, the relation between $E$ and $N$ is obtained and shown on
Fig.\eqref{fig:final_result_eigenvalue1}. The grey curves are obtained
by the WKB approximation from \prettyref{eq:leading_order_eigenvalue_result_Fake}.
The other colored curves are numerical results, which are displayed
in blue for $N\geq2$, in orange for $N<2$, and in green whenever
$N$ is near to a location of degeneracy. On Fig.\eqref{fig:final_result_eigenvalue1}
the magnified region where the degeneracy occurs clearly shows the
green data points, two of which are very close to the actual value
of the degenerated eigenvalue. We summarize all these green points
on the Tab.\eqref{tab:numerical_degeneracy}.

The leading-order WKB method is a very good approximation, since on
the most part of Fig.\eqref{fig:final_result_eigenvalue1} those grey
curves are covered by the blue and orange curves so that we barely
see them. However, whenever $N$ is approaching a location of degeneracy,
the WKB approximation is no longer reliable. The WKB also fails when
$N$ approaches to $1$, where only the ground state has real eigenvalue.
When $N=1$ exactly, we did not find any real eigenvalue, including
the ground level. We will talk more about the WKB approximation later.

\begin{table}[H]
\caption{\label{tab:numerical_degeneracy}Locations of the green data points
(two nearest to the degenerated eigenvalues).}

\centering{}%
\begin{tabular}{|c|c|c|}
\hline 
 & $N$ & Re$\left(E\right)$\tabularnewline
\hline 
\hline 
\multirow{2}{*}{Level 0} & 1.42210 & 3.798097503566341\tabularnewline
\cline{2-3} 
 & 1.42210 & 3.769947569720313\tabularnewline
\hline 
\multirow{2}{*}{Level 1} & 1.57145  & 6.931062951894809\tabularnewline
\cline{2-3} 
 & 1.57145 & 6.909904226441585\tabularnewline
\hline 
\multirow{2}{*}{Level 2} & 1.64860 &  10.19710564838468\tabularnewline
\cline{2-3} 
 & 1.64860 &  10.16647647154904\tabularnewline
\hline 
\multirow{2}{*}{Level 3} & 1.69810 & 13.56278552311201\tabularnewline
\cline{2-3} 
 & 1.69810 & 13.50221738682984\tabularnewline
\hline 
\multirow{2}{*}{Level 4} & 1.73330 & 16.98347623074032\tabularnewline
\cline{2-3} 
 & 1.73330 & 16.91803839426446\tabularnewline
\hline 
\multirow{2}{*}{Level 5} & 1.76000 & 20.46649448240978\tabularnewline
\cline{2-3} 
 & 1.76000 & 20.37974449784742\tabularnewline
\hline 
\end{tabular}
\end{table}

\begin{figure}[H]
\caption{\label{fig:final_result_eigenvalue2}Energy spectrum of the $2$nd
family from the pair of the green wedges (The grey curves are the
WKB approximation).}

\centering{}\includegraphics[scale=1.37]{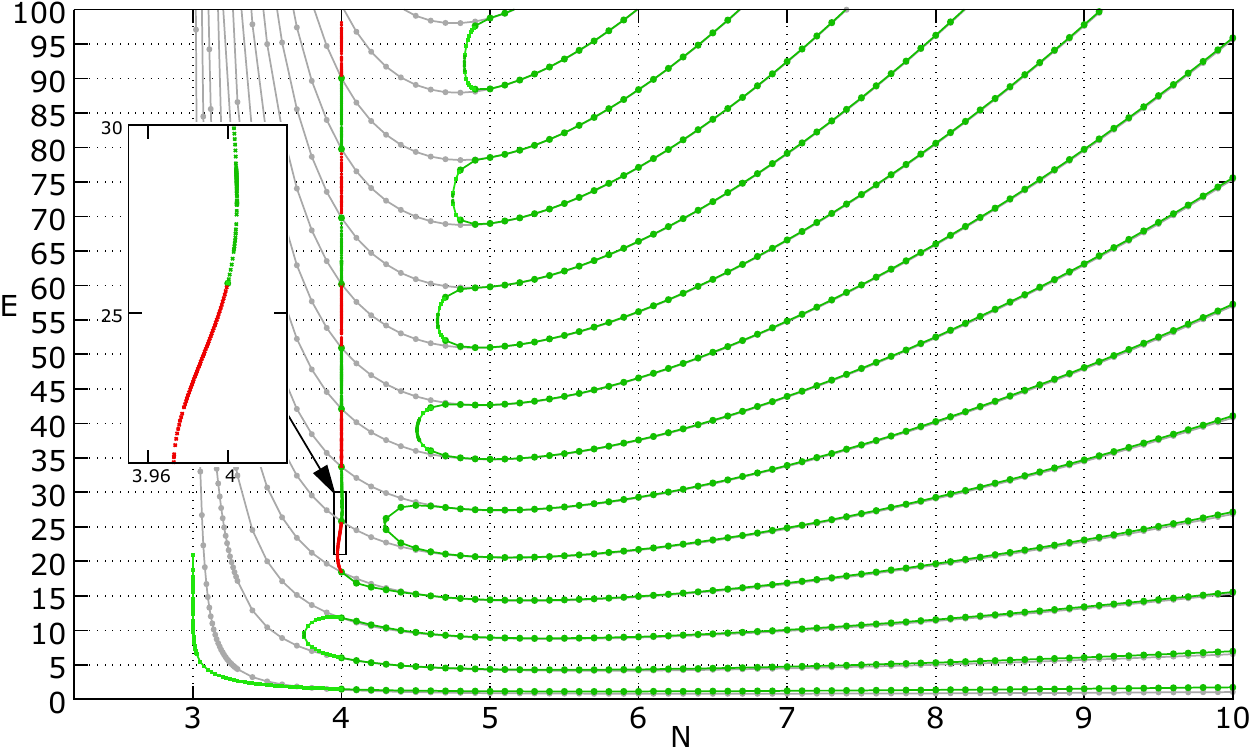}
\end{figure}

By the same way, on Fig.\eqref{fig:final_result_eigenvalue2} we obtain
the $2$nd family of eigenvalues from the green wedges defined by
Fig.\eqref{fig:many_wedges_multiplot}. The most interesting discovery
is that when $N$ is around $4$ and $E$ is about $20$, the eigenvalue
curve starts to go in vertical direction with horizontal oscillation,
whose amplitude decreases as $E$ increases. When this curve oscillates
to the left so that $N<4$, we use red color to plot the curve; when
oscillates to the right so that $N>4$, we use green color. When $E$
is around $20$, the curve is in red color and inbetween $3.97<N<4$;
when $E$ is above $90$, the curve is confined within $3.99999999998<N<4$,
which is on the top of the figure the red part of the curve, whose
oscillation is too small and can be almost ignored. We guess that
as $E$ further increases, the oscillation eventually breaks the limit
so that no device is able to detect such tiny oscillation, which generates
an illusion that the energy spectrum at $N=4$ is continuous for high
level (similar to the classical regime), rather than quantized. However,
theoretically, when $N$ is exactly equal to $4$, no matter how high
the energy $E$ is, the eigenvalues are still quantized and discrete
points whose locations distinguish the red and green part of the curve. 

Since this eigenvalue curve has infinite number of degeneracies so
we name it as a ``Curve with Infinite Number of Degeneracies'' or
``CIND''. Within the pair of the green wedges, how many CINDs are
there? Here is our conjecture without solid proof. Numerical result
shows that when $E$ is above $180$, the eigenvalue curve for $N=5$
becomes another CIND. However, we did not find such trend for the
eigenvalue curve when $N=6$, possibly because the entire region where
$N\geq6$ has unbroken $PT$-symmetry. The broken $PT$-symmetry happen
within the three regions where $5<N<6$, $4<N<5$ and $3<N<4$. We
observe that because of the existence of CINDs, $PT$-symmetry is
never broken when $N>3$ and $N$ is an integer. So we conclude that
CIND may only exist when $N$ is an integer, and there are two CINDs
associated with the pair of the green wedges.

\begin{figure}[H]
\caption{\label{fig:final_result_single_eigenvalue_curve}A single eigenvalue
curve from the pair of the green wedges.}

\centering{}\includegraphics[scale=1.02]{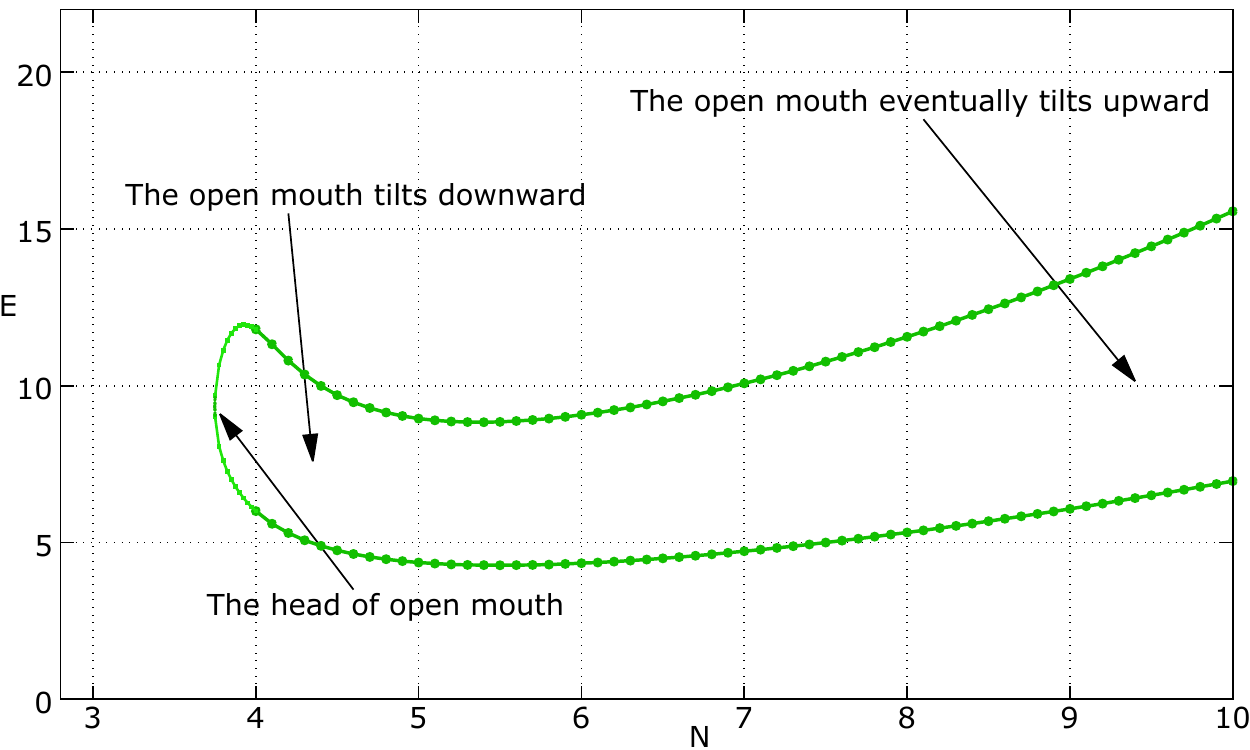}
\end{figure}

Before going further, we introduce the concept of open mouth. Most
eigenvalue curves have a standard shape similar to Fig.\eqref{fig:final_result_single_eigenvalue_curve},
where we call the empty region between two adjacent and connected
levels as an ``open mouth''. The head of an open mouth is the location
of the degeneracy. As $N$ increases, only two cases are observed:
one is that the open mouth tilts upward for increasing $N$, and the
other one is shown on Fig.\eqref{fig:final_result_single_eigenvalue_curve}
in which the open mouth initially tilts downward but eventually tilts
upward if $N$ is large enough. Particularly, for the green wedges
$N$ has to be around or larger than $6$ for that part of the open
mouth tilts upward. If the head of the open mouth is located far less
than $N=6$, then the tendency to initially tilt downward is more
pronounced. This is especially obvious for those low-lying states
as shown on Fig.\eqref{fig:final_result_single_eigenvalue_curve}
where the head of the open mouth is located at the region where $N<4$
and far less than $6$. 

\begin{figure}[H]
\caption{\label{fig:final_result_eigenvalue3}Energy spectrum of the $3$rd
family from the pair of the pink wedges (The grey curves are the WKB
approximation).}

\centering{}\includegraphics[scale=1.37]{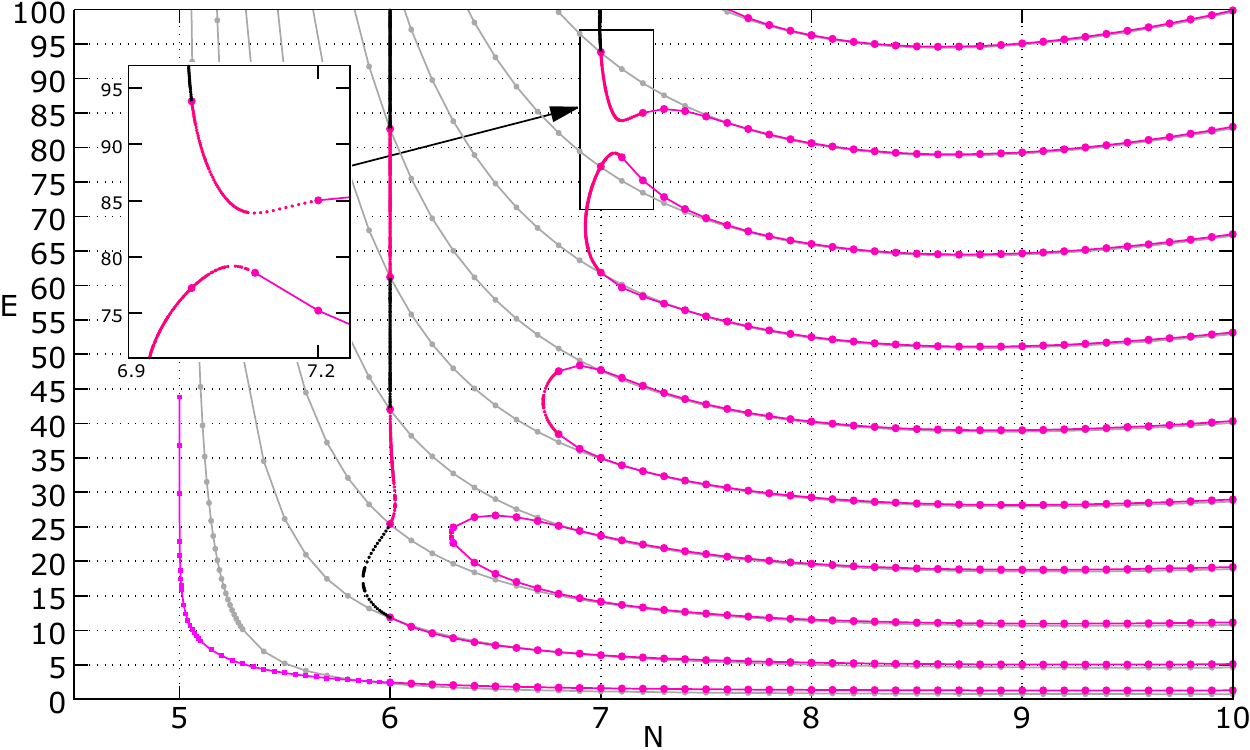}
\end{figure}

Why does the vertically straight line around $N=6$ differentiate
the behavior of the open mouth? When $N<6$, Fig.\eqref{fig:many_wedges_multiplot}
shows that the pair of the two green wedges moves above the real axis
so that the $PT$-symmetry of the wedges is broken for any non-integer
$N$, and this movement may cause the open mouth to tilt downward.
When $N>6$, Fig.\eqref{fig:many_wedges_multiplot} shows that the
green wedges moves below the real axis so that the $PT$-symmetry
is unbroken for any real $N$, and consequently the open mouth may
tilt upward. In short, the location of the $PT$-symmetric wedges
may determine which direction the open mouth tilts in. This also implies
that the open mouths of eigenvalue curves from the pink wedges (See
Fig.\eqref{fig:final_result_eigenvalue3}) will eventually tilt upward
as $N>10$. 

On Fig.\eqref{fig:final_result_eigenvalue3} we plot the $3$rd family
of eigenvalues for the pair of the pink wedges defined by Fig.\eqref{fig:many_wedges_multiplot}.
There are four CINDs, where CINDs for $N=8$ and $N=9$ happen in
much higher level. The region where $N\geq10$ has unbroken $PT$-symmetry,
whereas broken $PT$-symmetry happens within the region $N<10$ except
for integer $N$. As the previous case of the $2$nd family, the amplitude
of horizontal oscillation for CIND decreases as $E$ increases. Taking
the CIND for $N=6$ as an example, the part of the CIND where $N<6$
is plotted in black color and $N>6$ in pink color. Within the region
around $E=15$ and $N=6$, the black part of the CIND is confined
within $5.87<N<6$; whereas for the region where $E>90$, the CIND
(in black color again) is within $5.9999998<N<6$.

On Fig.\eqref{fig:final_result_eigenvalue3}, we magnify the most
interesting region, where two levels fail to connect and form a degeneracy
around $E=80$ because they are too close to the integer $N=7$, where
$PT$-symmetry is never broken. Consequently, the curve from the lower
level merges with the curve from the even lower level; whereas the
curve from the upper level form a CIND for $N=7$. This kind of behavior
is somewhat similar to the cohesion of liquid, where similar or identical
particles tend to cling to one another if the distance between them
is small enough. 

\begin{figure}[H]
\caption{\label{fig:final_result_eigenvalue_all_wedges}Four energy spectra
of the first four families from four pairs of $PT$-symmetric (orange,
green, pink, yellow) wedges.}

\centering{}\includegraphics[scale=1.37]{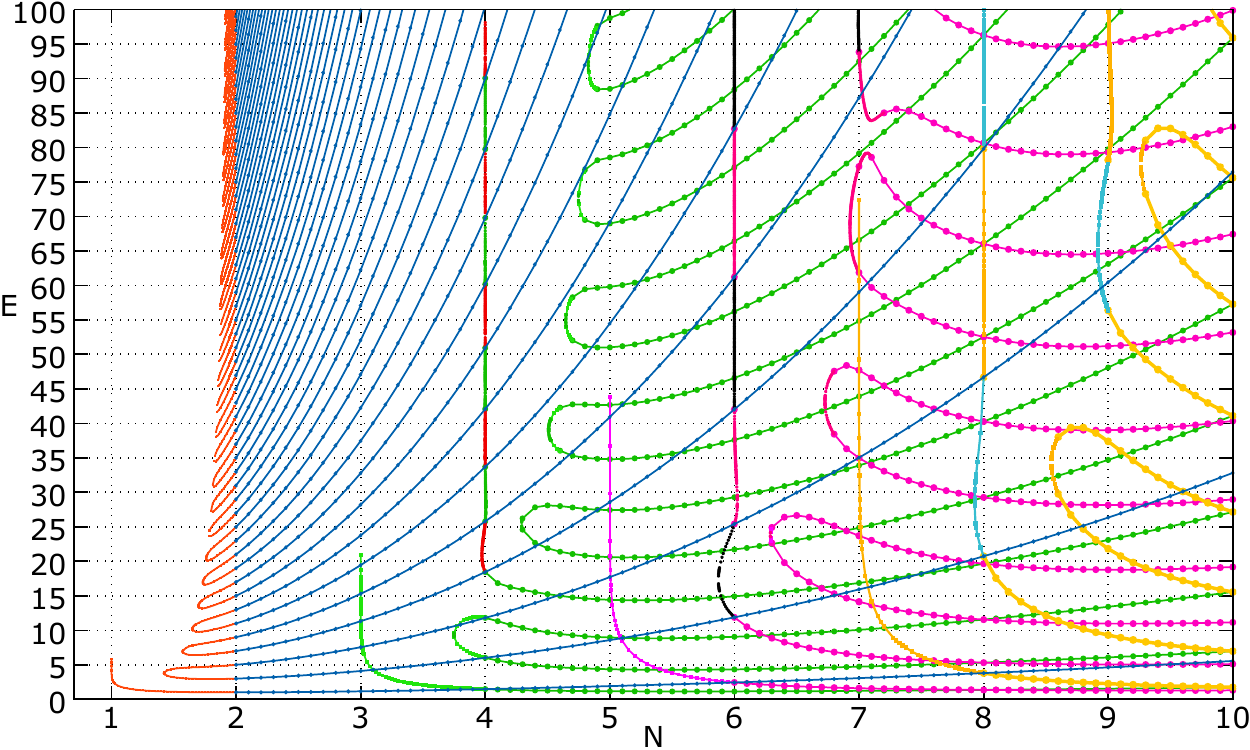}
\end{figure}

By including real eigenvalues from the four $PT$-symmetric wedges
(the orange, green, pink and yellow wedges), we plot the first four
families of energy spectra altogether shown on Fig.\eqref{fig:final_result_eigenvalue_all_wedges}.
An interesting feature is immediately spotted. All open mouths from
the $1$st family from the orange wedges tilt upward, while the open
mouths from the $3$rd family from the pink wedges tilt downward,
and the $4$th family tilts downward even more. This is because that
all heads of open mouths from the $1$st family including the low-lying
states are quite near to the vertically straight line $N=2$ where
differentiates the regions with broken and unbroken $PT$-symmetry
for the $1$st family, while heads of open mouths from the $4$th
family are quite far away from $N=14$ which differentiates the regions
with broken and unbroken $PT$-symmetry for the $4$th family.

Now let's talk about another interesting feature. At exactly $N=4$,
the eigenvalues from the $1$st and $2$nd family are equal, which
means that rather than two families, there is only one family of eigenvalues
at $N=4$. Similarly, at exactly $N=6$, the eigenvalues from the
$1$st and $3$rd family are equal, which means that rather than three
families, there are only two families at $N=6$. The most interesting
part is at exactly $N=8$ that not only the eigenvalues from the $1$st
and $4$th family are equal, but the $2$nd and $3$rd family are
also equal, so that rather than four families, there are only two
families at $N=8$. Do we expect this feature? Yes, because of symmetry.
When $N=4$, Fig.\eqref{fig:many_wedges_multiplot} shows that orange
and green wedges are symmetric with respect to the real axis. When
$N=6$, the orange and pink wedges are symmetric with respect to the
real axis while the pair of green wedges lies right on the real axis.
When $N=8$, not only the orange and yellow wedges but also the green
and pink wedges are symmetric with respect to the real axis. This
symmetry reduces the number of families of eigenvalues at even $N$
except when $N=2$. The moment when $N$ is an odd integer, a new
$PT$-symmetric wedges are born from the positive-imaginary axis and
consequently, a new family of spectrum is born.

\subsection{A comment on the WKB approximation}

By observing on previous figures, the leading-order WKB method from
\prettyref{eq:leading_order_eigenvalue_result_Fake} is a pretty good
approximation in those regions with unbroken $PT$-symmetry, including
those integers $N$ associated with CINDs. However, the WKB approximation
may fail wherever the $PT$-symmetry is broken. The reason why it
fails can be subtler than the reason provided by the paper\cite{Real_Spectra},
where it says that when $N<2$, the path along which the integral
$\intop_{x_{-}}^{x_{+}}dx\sqrt{E-V\left(x\right)}$ is real is in
the upper-half $x$ plane so that it crosses the cut on the positive-imaginary
axis and thus is not a continuous path joining the turning points.
This reason is only true if the pair of orange wedges from Fig.\eqref{fig:many_wedges_multiplot}
is chosen. Let's see why.

\begin{figure}[H]
\caption{\label{fig:N=00003D5_stoke_diagram-2}Stokes diagram for $N=5$ and
$E=1$, where the yellow points are the turning points, the green
dot-lines are Stokes lines and the orange dot-lines are anti-Stokes
lines.}

\centering{}\includegraphics[scale=1.5]{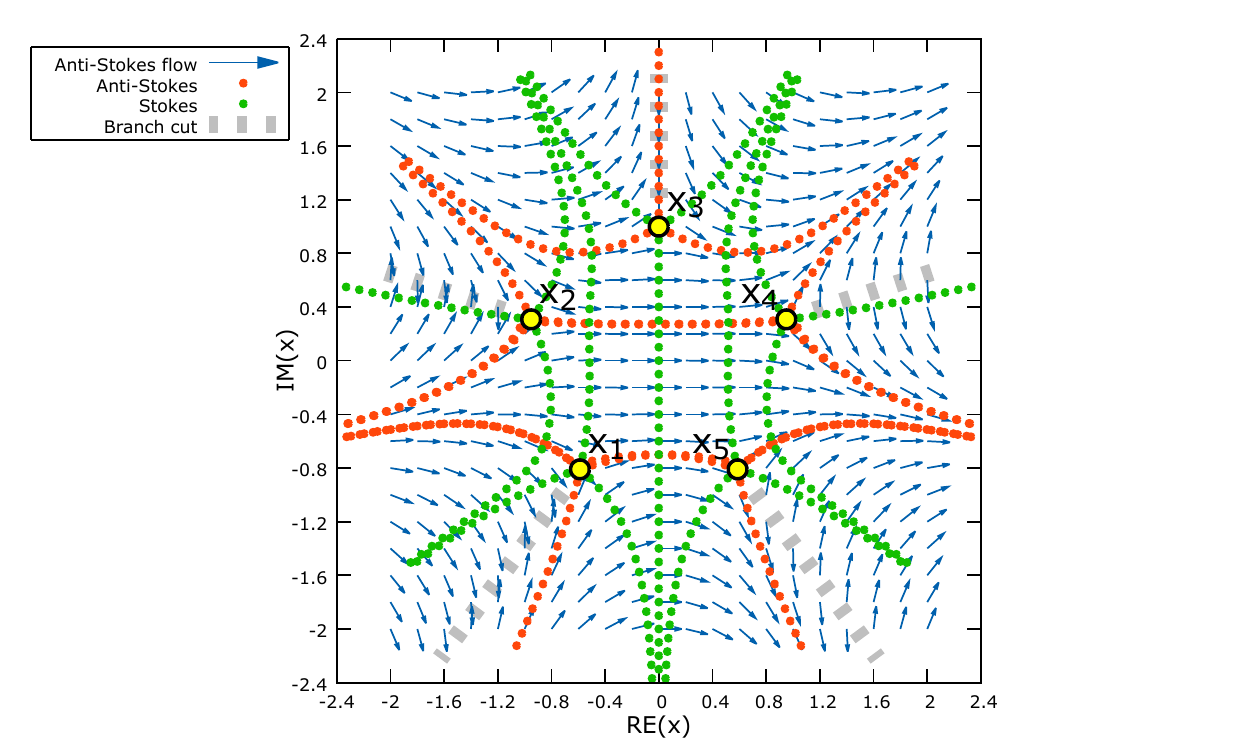}
\end{figure}

For the following $1$D Schrodinger equation, 
\[
-\frac{d^{2}\psi\left(x\right)}{dx^{2}}+V\left(x\right)\psi\left(x\right)=E\psi\left(x\right)\text{,}
\]
the leading-order WKB approximation gives two asymptotic solutions
as $\left|x\right|\rightarrow\infty$ 
\begin{equation}
\psi_{\pm}\left(x\right)\sim\frac{1}{Q\left(x\right)^{\nicefrac{1}{4}}}\exp\left(\pm i\intop_{x_{0}}^{x}\left[Q\left(t\right)\right]^{\nicefrac{1}{2}}dt\right)\qquad\mbox{where }Q\left(x\right)=E-V\left(x\right)=E+\left(ix\right)^{N}\text{,}\label{eq:complex_1st_order_sol}
\end{equation}
where $x_{0}$ is a turning point, and $x$ is a complex variable.
To be consistent with our previous definition, we define Stokes line
as 
\begin{equation}
\operatorname{Re}\left\{ \intop_{x_{0}}^{x}\left[Q\left(t\right)\right]^{\nicefrac{1}{2}}dt\right\} =0\text{,}\label{eq:def_Stokes_line}
\end{equation}
and anti-Stokes line as 
\[
\operatorname{Im}\left\{ \intop_{x_{0}}^{x}\left[Q\left(t\right)\right]^{\nicefrac{1}{2}}dt\right\} =0\text{,}
\]
Fig.\eqref{fig:N=00003D5_stoke_diagram-2} is called Stokes diagram
which shows the first-order approximation of the Stokes and anti-Stokes
lines when $N=5$. This diagram includes more detailed Stokes structure
than on Fig.\eqref{fig:indepdent_paths_N=00003D5}, where only the
Stokes structure at $\left|x\right|\rightarrow\infty$ is shown. (See
\cite[p.75]{Dr_white} to know how to generate Stokes diagram.) 

Fig.\eqref{fig:final_result_eigenvalue2} shows that the WKB method
from \prettyref{eq:leading_order_eigenvalue_result_Fake} is still
pretty good approximation to the eigenvalue within the green wedges
defined by Fig.\eqref{fig:indepdent_paths_N=00003D5} when $N=5$.
But Fig.\eqref{fig:N=00003D5_stoke_diagram-2} shows that the two
turning points $x_{2}$ and $x_{4}$ within the green wedges are joined
by an anti-Stokes line segment, which crosses the positive-imaginary
axis and by its definition implies that the integral $\intop_{x_{2}}^{x_{4}}dx\sqrt{E-V\left(x\right)}$
is real. So the argument provided by the paper\cite{Real_Spectra}
only works for non-integer $N$. For any integer $N$ associated with
a CIND, the $PT$-symmetry is not really broken for the given wedges. 

\section{Conclusion}

By using numerical and WKB approximation, we have answered the three
questions we posed in the abstract of this paper. Although these answers
lack of mathematical rigor and consequently can not guarantee to be
$100\%$ correct, we present students and researchers with wealthy information,
user-friendly interface and keen insight which are very useful under
the background of contemporary physics to understand the concept of
Stokes wedge and eigenvalue problem in complex plane. We believe that
a good understanding means not only understanding through rigorous
mathematical proof but also understanding through empirical evidence,
visualization and approximation. Those empirical observations are
essential for students to develop mathematical intuition and finally
come up with their own proofs. 

\bigskip{}

\textit{Acknowledgments} CT receives partially financial support from
NSERC Discovery Grant at Simon Fraser University in Canada.

\bibliographystyle{plain}
\nocite{*}
\bibliography{Paper_Tang2017.bbl}

\begin{thebibliography}{10}

\bibitem{Introduction_to_PT-Symmetric_Quantum_Theory}
C.~M. Bender.
\newblock {Introduction to {$\textit{PT}$}-Symmetric Quantum Theory}.
\newblock {\em Contemp.Phys.}, 46:277--292, 2005.

\bibitem{Make_sense}
C.~M. Bender.
\newblock {Making Sense of Non-Hermitian Hamiltonians}.
\newblock {\em Rept. Prog. Phys.}, 70:947, 2007.

\bibitem{complexWKB_bender}
C.~M. Bender, M.~Berry, P.~N. Meisinger, V.~M. Savage, and M.~Simsek.
\newblock {Complex WKB Analysis of Energy-Level Degeneracies of Non-Hermitian
  Hamiltonians}.
\newblock {\em Journal of Physics A: Mathematical and General}, 34:L31L36,
  2001.

\bibitem{Real_Spectra}
C.~M. Bender and S.~Boettcher.
\newblock {Real Spectra in Non-Hermitian Hamiltonians Having {$\textit{PT}$}
  Symmetry}.
\newblock {\em Phys. Rev. Lett.}, 80:5243--5246, Jun 1998.

\bibitem{bender2002two}
C.~M. Bender, S.~Boettcher, P.~N. Meisinger, and Q.~Wang.
\newblock {Two-point Green's Function in {$\textit{PT}$}-symmetric Theories}.
\newblock {\em Physics Letters A}, 302(5):286--290, 2002.

\bibitem{complex_extension_QM}
C.~M. Bender, D.~C. Brody, and H.~F. Jones.
\newblock {Complex Extension of Quantum Mechanics}.
\newblock {\em Phys. Rev. Lett.}, 89:270401, Dec 2002.
\newblock [Erratum: Phys. Rev. Lett.92,119902(2004)].

\bibitem{bender1999advanced}
C.~M. Bender and S.~A. Orszag.
\newblock {\em {Advanced Mathematical Methods for Scientists and Engineers I:
  Asymptotic Methods and Perturbation Theory}}.
\newblock Springer, 1999.

\bibitem{bender1993analytic}
C.~M. Bender and A.~Turbiner.
\newblock {Analytic Continuation of Eigenvalue Problems}.
\newblock {\em Physics Letters A}, 173(6):442--446, 1993.

\bibitem{dorey2001spectral}
P.~Dorey, C.~Dunning, and R.~Tateo.
\newblock {Spectral equivalences, Bethe Ansatz equations, and reality
  properties in {$\textit{PT}$}-symmetric quantum mechanics}.
\newblock {\em Journal of Physics A: Mathematical and General},
  34(28):5679--5704, 2001.

\bibitem{Gauss_legendre}
A.~Iserles.
\newblock {\em A First Course in the Numerical Analysis of Differential
  Equations}.
\newblock Cambridge University Press, New York, NY, USA, 2nd edition, 2008.

\bibitem{schmidt2013generation}
S.~Schmidt and S.~P. Klevansky.
\newblock {Generation of families of spectra in {$\textit{PT}$}-symmetric
  quantum mechanics and scalar bosonic field theory}.
\newblock {\em Philosophical Transactions of the Royal Society of London A:
  Mathematical, Physical and Engineering Sciences}, 371(1989):20120049, 2013.

\bibitem{complex_WKB_marker}
M.~Sorrell.
\newblock {Complex WKB Analysis of a {$\textit{PT}$}-symmetric Eigenvalue
  Problem}.
\newblock {\em Journal of Physics A Mathematical General}, 40:10319--10335, aug
  2007.

\bibitem{Dr_white}
R.~White.
\newblock {\em {Asymptotic Analysis of Differential Equations}}.
\newblock Imperial College Press, revised edition, 2012.

\end{thebibliography}

\end{document}